\documentclass[a4paper, aps, pra, 11pt, onecolumn, tightenlines, superscriptaddress]{revtex4-2}
\pdfoutput=1
\usepackage{preamble}

\usepackage{tikz}

\makeatletter
\@ifundefined{lem} {\newtheorem{lem}{Lemma}}{}
\@ifundefined{them}{}{}
\makeatother
\providecommand{\orcid}[1]{\href{https://orcid.org/#1}{%
  \IfFileExists{orcid.pdf}{\includegraphics[width=8pt]{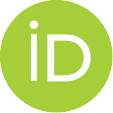}}%
                          {\textsuperscript{\tiny\textsc{iD}}}}}

\begin{document}
\title{Lattices, Gates, and Curves: GKP codes as a Rosetta stone}
\author{Jonathan Conrad~\orcid{0000-0001-6120-9930}}
\thanks{\url{jonathan.conrad@epfl.ch}}
\affiliation{\'Ecole Polytechnique F\'ed\'erale de Lausanne (EPFL), Quantum complexity and cryptography laboratory, Station 15, 1015 Lausanne}
\affiliation{Dahlem Center for Complex Quantum Systems, Physics Department, Freie
Universit{\"a}t Berlin, Arnimallee 14, 14195 Berlin, Germany}
\affiliation{Helmholtz-Zentrum Berlin f{\"u}r Materialien und Energie, Hahn-Meitner-Platz 1, 14109
Berlin, Germany}
\author{Ansgar G.\ Burchards~\orcid{0000-0002-6172-7962}}
\affiliation{Dahlem Center for Complex Quantum Systems, Physics Department, Freie
Universit{\"a}t Berlin, Arnimallee 14, 14195 Berlin, Germany}
\author{Steven T.\ Flammia~\orcid{0000-0002-3975-0226}}
\affiliation{Department of Computer Science, Virginia Tech, Alexandria, USA}
\affiliation{Phasecraft Inc., Washington DC, USA}

\date{\today}

\begin{abstract}
We explain how GKP Clifford gates arise as symplectic automorphisms of the corresponding GKP lattice and show that, for scaled codes of type $D=dI_n$, their integral action agrees with the action of the mapping class group of a genus-$n$ surface on first homology. 
This correspondence introduces a topological interpretation of fault tolerance for GKP codes and motivates the connection between GKP codes (lattices), their Clifford gates, and algebraic curves, which we explore in depth. 
For a single-mode GKP code, we identify the space of oriented fixed-covolume lattice realizations with $S^3 - K$, the three sphere $S^3$ with a trefoil knot $K$ removed, and explain how logical degrees of freedom arise from the choice of a level structure on the corresponding curves. Specified Clifford implementations define loops in the space of lattices whose trefoil linking number is a topological invariant.
Finally, we relate our observations to the idea of fiber bundle fault tolerance as proposed by Gottesman and Zhang \cite{gottesman2017fibre} for the GKP code, where logical Clifford and Pauli operations of the single-mode GKP code arise as the monodromy representation of a finite covering of the space of nonzero-distance GKP codes. 
\end{abstract}
\maketitle

\section{Introduction}

\begin{figure}
\includegraphics[width=.3\columnwidth]{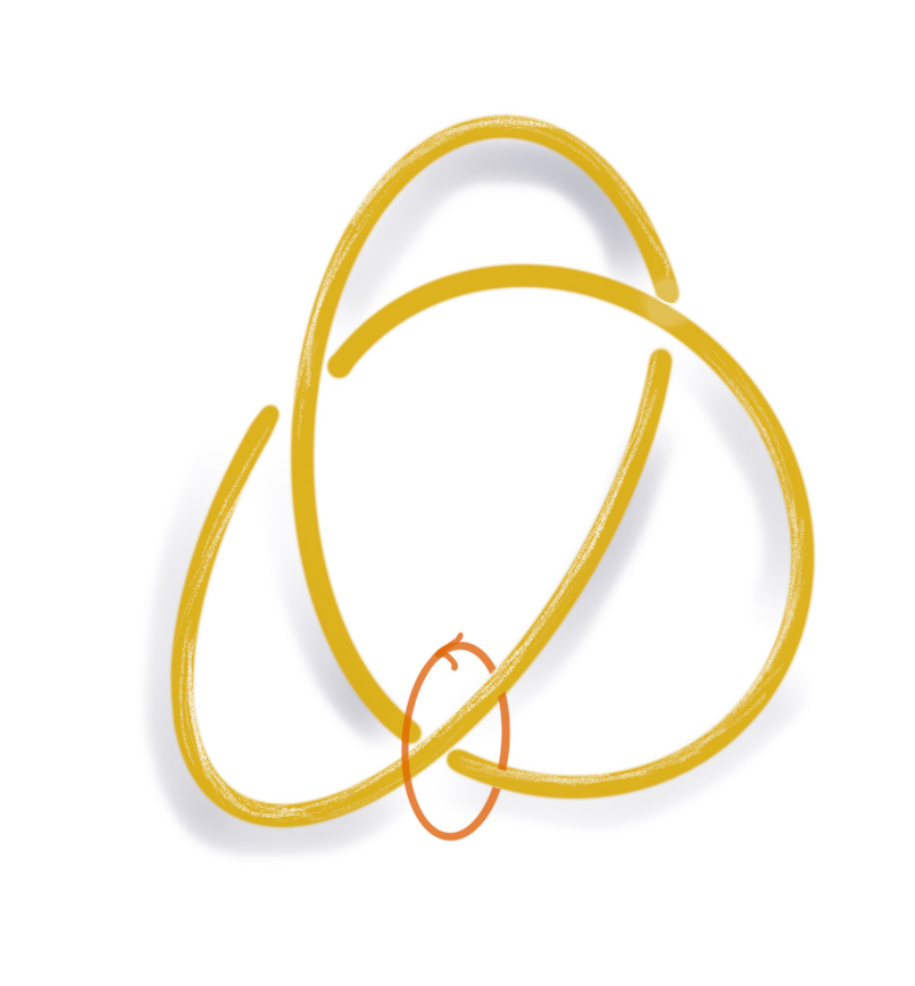}
\caption{The trefoil knot represents the degenerate zero-distance locus removed from the space of oriented fixed-covolume single-mode lattice realizations. We illustrate a non-trivial link with the trefoil knot given by a $\pi/2$ rotation of the square lattice $\Lambda \mapsto e^{i\phi}\Lambda,\; \phi \in \lrq{0, \pi/2}$. 
The square lattice corresponds to the standard GKP code, and this link implements a logical Hadamard gate.}\label{fig:trefoil}
\end{figure}

Gottesman-Kitaev-Preskill (GKP) codes \cite{GKP} are bosonic quantum error correcting codes that encode discrete information into a collection of quantum harmonic oscillators. 
The code states are defined by translation invariance in their phase space representation which carries the structure of a lattice $\CL$ so that small displacements of these states can be measured with high precision without collapsing the structure of the states \cite{Terhal_2020, Duivenvoorden_Sensor}. 
As quantum error correcting codes, GKP codes are promising candidates to protect against photon loss \cite{Albert_2018, Noh_Capacity, Terhal_2020}, and recent experiments \cite{Sivak_2023, Fluehmann_2019, Campagne_Ibarcq_2020} have demonstrated further support for their practical relevance. 

Aside from their promise as a quantum memory, GKP codes allow for the implementation of the logical Clifford group through only Gaussian unitary operations. 
It has also been observed that magic states of the GKP code can be obtained by performing error correction on the Gaussian vacuum state \cite{AllGaussian}. 
With the promise of fault-tolerant logical Clifford gates, GKP codes are hence expected to present a viable platform for large-scale fault-tolerant quantum computation once stabilizer measurements or state preparation can be performed with high precision. 

As we have not formally defined the property of \textit{fault-tolerance} in the preceding paragraph, we note that in the literature the precise meaning of this property is formulated depending the noise model, computational setting, research themes and target platforms and varying level of targeted rigour. Intuitively, \textit{fault-tolerance} should be taken verbatim: the implementation of a process (say a quantum algorithm with classical or quantum inputs and classical or quantum outputs), whose macroscopic behaviour is robust to faults in the precise microscopic implementation.

Though the focus in this paper is on GKP codes, we note that a satisfactory general definition of fault tolerance as a microscopic and testable property is a present problem already for qubit codes~\cite{GottesmanQECtalk, gottesman2022opportunities}. With this work we attempt to contribute to finding a more concrete characterization by showing that the concept of \emph{fault-tolerance} for unitary transformations of the GKP code can be understood as a smooth curve through the set of positive-distance GKP codes. 

One class of codes where fault tolerance has a clear \emph{topological} interpretation is the class of homological codes~\cite{Kitaev_2003}. 
In homological codes, certain logical gates are operators whose support lies on homologically nontrivial loops on the physical qubit geometry and thus are inherently robust against local noise processes. 

A general topological approach to fault tolerance has been proposed by Gottesman and Zhang \cite{gottesman2017fibre}, where fault-tolerant implementations of logical gates are proposed to be identified with non-contractible loops of a suitably chosen set of codes for a given physical Hilbert space. 
This makes concrete the interpretation of fault-tolerant logical gates as a sequence of smooth code-switching operations that preserve some global (e.g.\ distance) property. 
This proposal yields a \emph{topological} understanding of fault tolerance and has the potential to provide a definition that unifies different practice-driven approaches. 
A few examples of this approach to fault-tolerant protocols for qubit-based error correcting codes have been discussed in \cite{gottesman2017fibre}, including transversal gates, as well as the braiding of anyons in lattice models. 

Here we establish a first link between bosonic error correction and topological notions of fault tolerance by considering the topological properties of logical GKP Clifford gates and their implementation by smooth paths of Gaussian operations. 
We explore and exploit a direct connection between GKP codes via their associated lattices, Riemann surfaces, and algebraic curves; specifically, we identify single-mode lattice shapes modulo global rotation with elliptic curves; a level structure supplies additional logical-label data. 
We show how the understanding of logical Clifford gates for the GKP codes leads to a natural correspondence between smooth paths of lattice realizations and paths in the space of suitable algebraic curves.

This correspondence allows us to identify the space of oriented fixed-covolume single-mode lattice realizations, up to integral basis change, with $S^3-K$, a three sphere $S^3$ with a trefoil knot $K$ removed.
Closed loops on this space that link with this knot arise from implementations of fault-tolerant logical Clifford gates on the GKP code. 
Using tools from the theory of modular forms, we show how the ``knot defect" provided by the trefoil can be interpreted as the limit where GKP codes have zero distance, i.e., the phase space lattice becomes degenerate and the codes can no longer correct a general error. %
Conversely, any point away from the trefoil corresponds to a GKP code with nonzero distance. 
Consequently, paths which avoid the trefoil correspond to fault-tolerant sequences of code-switches between GKP codes, in the sense of Gottesman and Zhang. 
It is therefore natural to investigate closed paths with nontrivial homotopy. 

Finally, we construct the natural orbifold family of single-mode GKP code shapes with nonzero syndrome. Away from points with nontrivial symmetries, the quotient map is a covering whose monodromy records symplectic Clifford actions and Pauli displacement labels, providing a geometric counterpart to the covering-space structure of Ref.~\cite{gottesman2017fibre}.
In the present work, we focus on full-rank GKP codes encoding finite-dimensional logical systems.  The subsequently published Ref.~\cite{Burchards_GeometricGKP} develops the general multimode GKP moduli space and realizes the vision of Gottesman and Zhang for logical GKP Clifford operations via a covering space analysis; the present paper supplies the initial single-mode algebraic-geometric and modular-knot analysis.
\medskip
\\
Our paper is organized as follows. 
In section \ref{sec:Cliffords} we provide a brief introduction to GKP codes and discuss their Clifford groups; for scaled codes, the integral symplectic action agrees with the mapping-class-group action on first homology. 
We elaborate on the source of this connection and explain how GKP lattices in the Jacobian locus arise from compact Riemann surfaces which correspond to algebraic curves.
Exploring this connection more deeply, we discuss how single-mode GKP codes can be understood as instances of complex elliptic curves with a level structure so that, in section \ref{sec:GKP_FT}, we build on their well-explored moduli to construct the corresponding finite quotient for GKP codes. 
This construction explains the relationship between the space of GKP codes and the underlying topology of the oriented fixed-covolume single-mode lattice-realization space and, finally, we compare the covering obtained away from points with nontrivial symmetries with structure needed for fiber bundle fault tolerance by Gottesman and Zhang. 
We close with a discussion and point to possible lines of research building on our exposition.

\section{The GKP code and its Cliffords}\label{sec:Cliffords}

We develop the mathematical structure of GKP codes to understand the structure of their Clifford gates and their fault tolerance. 
A lattice-theoretic treatment of GKP codes can be found in \cite{Conrad_2022, Royer_2022, Schmidt_2022, HarringtonPreskill, Harrington_Thesis, GKP} and more details regarding displacement and Gaussian unitary operators are discussed in appendix \ref{app:QHO}, which we refer to for further background.
Here a GKP code is the simultaneous $+1$ eigenspace of an abelian group of displacement operators. To avoid discussing the physical structure of GKP states (which is a separate intricate topic~\cite{GKP}), we place our focus on describing the groups of stabilizer and logical operators associated to GKP codes.

GKP codes are described by a full-rank \textit{symplectically integral}\footnote{Also called \textit{weakly symplectically self-dual} in \cite{conrad2023good}.} lattice $\CL\subseteq\CL^{\perp}\subset \R^{2n}$, where the symplectic dual is defined by $\CL^{\perp}:=\{\bs{\eta}:\bs{\eta}^{T}J\bs{\xi}\in\Z\text{ for every }\bs{\xi}\in\CL\}$; thus symplectic integrality means that $\CL\subseteq\CL^{\perp}$. We use the convention $J=\begin{pmatrix}
0 & I_n \\ -I_n & 0
\end{pmatrix}$. 
The lattice $\CL$ describes the stabilizer group $\mathcal{S}$ of the GKP code which is, up to a phase factor \cite{Conrad_2022}, given by displacements 
\begin{equation}
    D\lr{\bs{\xi}} \coloneqq e^{-i \sqrt{2\pi} \bs{\xi}^T J \bs{\hat{x}}},\quad D(\bs{\xi})D(\bs{\eta})=e^{-i2\pi\bs{\xi}^{T}J\bs{\eta}}D(\bs{\eta})D(\bs{\xi}), \label{eq:displConvJ}
\end{equation} 
acting on $n$ modes of a generalized quantum harmonic oscillator, with $\bs{\hat{x}}=\lr{\hat{x}_1,\hdots, \hat{x}_n,\hat{p}_1,\hdots, \hat{p}_n}^T$ being the vector of quadrature operators (we set $\hbar=1$). 
Displacements with amplitudes in the dual lattice $\CL^{\perp}$ describe, via the same lift, eq.~\eqref{eq:displConvJ}, all displacement operators that commute with the stabilizer group and these are hence identified with the centralizer $\mathcal{C}\lr{\mathcal{S}}$ within the set of displacement operators. 
The displacements associated to representatives of (generalized) logical Pauli operators, which are also known as Heisenberg-Weyl operators, are given by $\mathcal{C}\lr{\mathcal{S}} / \mathcal{S}$ and are hence provided by the lattice cosets $\CL^{\perp}/\CL$. 

We describe lattices $\CL={\rm span}_{\Z}\lr{M}, \CL^{\perp}={\rm span}_{\Z}\lr{M^{\perp}}$ by the integer row-span of corresponding generator matrices $M, M^{\perp} \in \R^{2n \times 2n}$. 
The condition for the GKP stabilizer group to be abelian, $\CL\subseteq \CL^{\perp}$ is equivalent to the integrality of the symplectic Gram matrix $A\coloneqq MJM^T$ and bases can always be found such that $M=AM^{\perp}$ by means of a basis transformation. 

The entries of $A$ are exactly the symplectic pairings of the lattice basis vectors, so they are integral if and only if $\CL\subseteq\CL^{\perp}$. It holds that $|\CL^{\perp}/\CL|=|\det(MJM^T)|=\operatorname{vol}(\CL)^2=d^2$, where $d=\prod_i d_i$ is the dimension of the encoded Hilbert space.
All elements in the orbit $\GL_{2n}\lr{\Z}M$ defined via left action describe the same lattice $\CL$ (and similarly for $M^{\perp}$) \footnote{In fact the orbit $\GL_{2n}\lr{\Z}M$ describes the set of all lattice bases that generate the same lattice.}.
We choose to work more conveniently with basis transformations that preserve the sign of the volume form $\det \lr{\CL}=\det \lr{M}$ without loss of generality.
For symplectically integral lattices, a canonical basis can always be found such that $A$ splits into a direct sum of $2 \times 2$ skew-symmetric blocks, $A=A_D=J_2\otimes D$ where $D={\rm diag} \lr{d_1,\hdots, d_n}$ and where the $d_j$ obey the divisibility condition $d_1 | d_2 | \hdots | d_n$. 
The invariant factors $d_1,\ldots,d_n$ are unique~\cite{Bourbaki9}, although the basis is not; we refer to $D$ as the \textit{type} of the lattice. For example, a qubit--qutrit product has canonical invariant factors $(1,6)$ rather than $(2,3)$.

In general, a lattice basis generating a GKP code of type $D$ can be obtained as a sublattice of a  symplectic lattice (for which $D=I_n$) that is generated by a symplectic matrix $M_0: M_0^TJM_0=J$ to
\begin{equation}
    M:=(D\oplus I) M_0. \label{eq:scaledGKP}
\end{equation}
Here and below $A\oplus B$ denotes the block-diagonal matrix ${\rm diag}(A,B)$. Equivalently, this way to construct a symplectically integral lattice of type $D$ can also be understood as the fact that any GKP code can be obtained by performing a symplectic transformation $S=M_0^T$ on a rectangular GKP code generated by $D\oplus I$ \cite{Conrad_2022}. %

Due to symplectic integrality \cite{Bourbaki9}, there always exists a basis, such that $M=(\bs{\xi}_1\,\hdots\, \bs{\xi}_{2n} )^T$ is given by symplectically conjugate pairs of vectors
\begin{align}
    \lr{\bs{\xi}_{i}, \bs{\xi}_{i+n}}:\, &\bs{\xi}_{i}^T J\bs{\xi}_{i+n}=d_i \;\;\forall i\in \lrc{1,\hdots, n}, \nonumber \\
    &\bs{\xi}_{i}^T J\bs{\xi}_{j}=0 \;\; \forall j\neq i+n\;\;\forall   i\in \lrc{1,\hdots, n},
\end{align}
and similarly, the dual basis $M^{\perp}=(\bs{\xi}_1^{\perp}\,\hdots\, \bs{\xi}_{2n}^{\perp} )^T$ can be arranged into pairs of vectors $(\bs{e}_i, \bs{f}_i)=(\bs{\xi}^{\perp}_{i}, \bs{\xi}^{\perp}_{i+n})$ with
\begin{align}
\bs{e}_i^T J\bs{f}_{i}&=\frac{1}{d_i} \;\; \forall i\in \lrc{1,\hdots, n} , \nonumber \\
 \bs{e}_i^T J\bs{f}_{j}&=0\;\; \forall  j\neq i\, \in \lrc{1,\hdots, n}.
\end{align}
These vectors constitute precisely the canonical representatives for the logical generalized Pauli operators of the GKP code and satisfy the rules of the desired Heisenberg-Weyl algebra
\begin{align}
\label{eq:HW_operators}
    X_i :=  D\lr{\bs{e}_i}, \, Z_i :=D\lr{\bs{f}_i}: \quad
    Z_iX_i=e^{i\frac{2\pi}{d_i}}X_iZ_i, \quad
    X_i^{d_i}, Z_i^{d_i} \in \CS, \quad
    [X_i, Z_j] =0\;\forall i\neq j\,.  
\end{align}

Finally, a distance 
\begin{equation}
    \Delta_{\rm GKP}\lr{\CL}\coloneqq\min_{\bs{x} \in \CL^{\perp}\setminus \CL} \|\bs{x}\|_2
\end{equation}
has been defined for GKP codes in \cite{Conrad_2022, GKP}, which captures its resilience against typical noise processes such as adversarial or stochastic Gaussian errors. 
See \cite{Conrad_2022,conrad2023good} for a more in-depth discussion of the distance.

Throughout this work, we will often focus on \emph{scaled GKP codes}, which are GKP codes obtained from $D=d I_n$ so that (up to a squeezing operation) the GKP lattice corresponds to a rescaled symplectically self-dual lattice $\CL=\sqrt{d}\CL_0$. 
Such scaled GKP codes can be understood to encode $n$ qudits with dimensions $d$ and the distance is given by the rescaled length of the shortest lattice vector of the symplectic lattice $\Delta_{\rm GKP}\lr{\CL}=d^{-\frac{1}{2}}\lambda_1\lr{\CL_0}$. 
The subsequent discussion is restricted to these scaled codes of common qudit dimension $d$; we do not assume that $d$ is prime.
 
\medskip
\paragraph*{Example: the square lattice, $\CL=\sqrt{2}\Z^2$.}
The simplest and most frequently discussed (scaled) GKP code is obtained by rescaling the symplectically self-dual lattice $\CL_0 = \Z^2$ with generator $M_0=I$ to a code that encodes one qubit into a quantum harmonic oscillator with $\CL=\sqrt{2}\Z^2$ and  $\CL^{\perp}=\sqrt{2}^{-1}\Z^2$. 
This code has distance $\Delta_{\rm GKP}=1/\sqrt{2}$ \cite{Conrad_2022}. 

\medskip
\paragraph*{Example: the hexagonal lattice, $\CL=\sqrt{2}A_2$.} 

The $A_2$ lattice, often also known as the hexagonal lattice, has generator matrix
\begin{equation}
M_{A_2}\coloneqq\frac{1}{\sqrt{2\sqrt{3}}}\begin{pmatrix}
2 & 0 \\ 1 & \sqrt{3}
\end{pmatrix},\label{eq:MA2}
\end{equation}
and the distance of the GKP code obtained by its $d=2$ scaling is $\Delta_{\rm GKP}=3^{-\frac{1}{4}}$. 
The hexagonal lattice implements the densest sphere packing in $2$ dimensions and hence also yields the single-mode GKP code with the largest distance.

\begin{figure}
\centering
\includegraphics[width=.6\columnwidth]{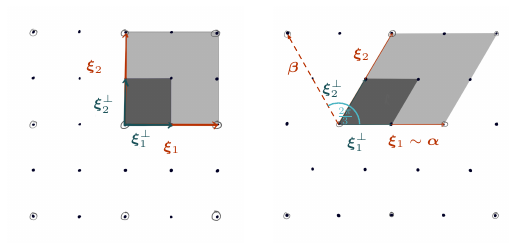}
\caption{
The symplectic lattices $\Z^2$ (left) and $A_2$ (right) scaled by $d=2$ and their respective (dual) unit cells. 
The logical displacement amplitudes are marked in turquoise and stabilizer displacements are marked in red.
}
\label{fig:Z2_A2}
\end{figure}

The set of GKP \emph{Clifford gates} forms a special and important set of gates that act on the logical space of a GKP code.
We describe GKP Clifford transformations at the level of lattice and displacement labels using the affine symplectic group
\begin{equation}
{\rm Cliff}\lr{D}\coloneqq \Aut_{\infty}^{S}\lr{\CL^{\perp}}=\Aut^S\lr{\CL^{\perp}} \ltimes \CL^{\perp},
\label{eq:affine-clifford-labels}
\end{equation}
Here $\Aut^S$ denotes linear symplectic lattice automorphisms and $\Aut^S_{\infty}$ the corresponding affine group. The semidirect product uses $S:\bs{\eta}\mapsto S\bs{\eta}$. Displacements change Paulis only by the phase in Eq.~\eqref{eq:displacement_commutator}; the linear part fixes zero and permutes logical Pauli labels.\footnote{The affine lattice group in Eq.~(\ref{eq:affine-clifford-labels}) is distinct from the finite logical Clifford group modulo overall phase. For a single qudit with $4\mid d$, the latter is a nonsplit extension of $\Sp_2(\Z_d)$ by $(\Z/d\Z)^2$, so its multiplication cannot be described by the corresponding semidirect-product law.}

In the remainder of this article, we focus on the symplectic representation of Clifford operators on logical Pauli labels $\CL^{\perp}/\CL$. For scaled codes, these labels form the symplectic module $(\Z_d)^{2n}$, a vector space when $d$ is prime. Logical Pauli displacement labels are retained separately in the affine constructions. We adopt the convention to call a Clifford gate trivial if its adjoint action on Pauli operators is projectively trivial.

Symplectic automorphisms transform vectors constituting the lattice basis $M$ in a way that only implements a symplectic change of basis while leaving the lattice as a geometric object invariant, 
\begin{equation}
\label{eq:AutS}
    \text{Aut}^{S}(\mathcal{L}) \coloneqq \{ g \in \Sp_{2n}\lr{\mathbb{R}} |\,  \exists U \in \SL_{2n}( \mathbb{Z}):  UM  = M g^{T}\},
\end{equation}
and we refer to the basis transformation $U\in \Sp^D_{2n}( \mathbb{Z})$ as the \textit{integral representation} of the corresponding element.
An important subgroup is the group of symplectic orthogonal automorphisms $\Aut^{SO}\lr{\mathcal{L}}=\Aut^{S}\lr{\mathcal{L}}\cap O_{2n}\lr{\R}$ which is significant due to its interpretation as GKP Clifford operations realizable via passive linear optics without squeezing~\cite{Weedbrook_2012}.

For GKP codes specified by a lattice $\CL \subseteq\CL^{\perp}$, each element of the symplectic automorphism group is uniquely specified by its integral representation given by the group
\begin{equation}
\label{eq:SpD}
    \Sp_{2n}^D\lr{\Z} \coloneqq \{U\in \GL_{2n}\lr{\Z}:\; UA_DU^T=A_D\}\,.
\end{equation}
These transformations preserve the symplectic form in its canonical basis $A_D=J_2\otimes D$ \cite{Birkenhake_2004}. 
Although Eq.~\eqref{eq:SpD} is written inside $\GL_{2n}(\Z)$, preservation of the symplectic form forces $\det U=1$.
We have the following.
\begin{lem}[Integral representation of linear GKP Cliffords]\label{prop:cliff}
Let $\CL$ be a full-rank symplectically integral lattice of type $D$, with basis matrix $M$ satisfying $MJM^T=A_D$. For each $g\in\Aut^S(\CL)$, let $U_g\in\GL_{2n}(\Z)$ be the unique matrix satisfying $U_gM=Mg^T$. Then $g\longmapsto U_g^{-1}$
is a group isomorphism
\begin{equation}
\Aut^S(\CL)\cong\Sp_{2n}^D(\Z)
=\{U\in\GL_{2n}(\Z):UA_DU^T=A_D\}.
\end{equation}
\end{lem}
\begin{proof}
Lemma~\ref{lem:integral_rep} gives the bijection between $g$ and $U_g$. Moreover, $U_g^{-1}U_h^{-1}M=Mg^{-T}h^{-T}=M(gh)^{-T}$, hence $g\mapsto U_g^{-1}$ is a homomorphism.
\end{proof}
Corollary~\ref{cor:dual-aut}, stated and proven in the appendix, additionally gives $\Aut^S(\CL)=\Aut^S(\CL^\perp)$.

The integral representation of the automorphisms in eq.~\eqref{eq:SpD} can be understood to represent the symplectic action of Clifford operators on the logical Heisenberg--Weyl labels in eq.~\eqref{eq:HW_operators}. 
If a logical Heisenberg-Weyl operator $O\lr{\bs{l}}=\prod_{i=1}^n X_i^{l_i} Z_i^{l_{i+n}}$ with $X_i, Z_i$ from eq.~\eqref{eq:HW_operators} is specified \footnote{That is, up to phases as we usually care about the action of these operators on a projective Hilbert space.} by a vector $\bs{l} \in \Z^{2n}_D$, we mean that labels in $\Z^{2n}$ are reduced component-wise modulo $D\oplus D$.
With the canonical dual basis $M^\perp=(D\oplus D)^{-1}M$, the relation $U_gM=Mg^T$ gives $M^\perp g^T=V_g M^\perp$ with $V_g=(D\oplus D)^{-1}U_g(D\oplus D)$. Thus a column vector of logical Pauli labels transforms as $\bs l\mapsto V_g^T\bs l\!\mod D\oplus D$.
The reduction modulo $D\oplus D$ therefore implements the equivalence relation given by the stabilizers as translations by elements in $\CL$ and corresponds to the projection $\Aut^S\lr{\CL^{\perp}} \rightarrow \Aut^S\lr{\CL^{\perp}/\CL}$. 
 
\begin{figure}
\centering
\includegraphics[width=.3\textwidth]{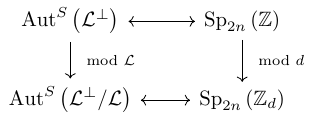}
\caption{Commutative diagram for the structure of nontrivial Cliffords for scaled GKP codes. Note that for $n=1$ we have the equality $\SL_{2}\lr{\mathbb{Z}}=\Sp_{2}\lr{\mathbb{Z}}$.}\label{fig:Cliffordn1}
\end{figure}

Henceforth we focus on scaled GKP codes; recall that for these we set $D=dI_n$, so that $\CL=d\CL^{\perp}$, which also implies $U_g=V_g \in \Sp_{2n}\lr{\Z}$ for the integral representations of the symplectic automorphisms.
The relationships between the integral and real representations of symplectic automorphisms for scaled GKP codes and their logical actions are illustrated in fig.~\ref{fig:Cliffordn1}. 
Reduction modulo $d$ gives $\Sp_{2n}(\Z)\to\Sp_{2n}(\Z_d)$ with kernel the principal congruence subgroup $\Gamma_{2n}(d)$; its induced action on $\CL^{\perp}/\CL$ is the usual symplectic representation of the qudit Clifford group.

The integral symplectic automorphism group of the self-dual lattice $\CL_0$ can be generated by \textit{symplectic transvections} with $\bs{\alpha}\in\CL_0$~\cite{omeara_symplectic_1978},
\begin{equation}
\label{eq:transvection}
    t_{\bs{\alpha}}\coloneqq I+\bs{\alpha}\bs{\alpha}^TJ\,,
\end{equation}
which are implemented via the Gaussian unitaries 
\begin{equation}
    U_{\bs{\alpha}}\coloneqq e^{-\frac{i}{2}\lr{\bs{\alpha}^TJ\bs{\hat{x}}}^2}
\end{equation}
that have squeezing value bounded by $\rm sq\lr{t_{\bs{\alpha}}}:=\|t_{\bs{\alpha}}\|_2\leq 1+\|\bs{\alpha}\|_2^2 $. 

The symplectic transvection maps $\bs{x}\mapsto\bs{x}+\bs{\alpha}(\bs{\alpha}^TJ\bs{x})$ and is therefore a shear; symplectic lattices $\CL_0$ are preserved under transvections by vectors in $\CL_0$.
For elements of a scaled GKP code $\CL^{\perp}=\sqrt{d}^{-1}\CL_0$, a symplectic transvection by one of the canonical basis vectors acts non-trivially on its partner,
\begin{equation}
    t_{\sqrt{d}\bs{e}_i} \bs{f}_i =\bs{f}_i + \bs{e}_i\,,
\end{equation}
and trivially on every other canonical basis vector. 
Using $t_{\bs{\alpha}}t_{\bs{\beta}}t^{-1}_{\bs{\alpha}}=t_{t_{\bs{\alpha}}(\bs{\beta})}$, one sees that $t_{\bs{\alpha}}$ and $t_{\bs{\beta}}$ commute when $\bs{\alpha}^TJ\bs{\beta}=0$.

In fact, symplectic transvections represent \textit{Dehn twists} on compact genus-$n$ surfaces $S_n$, while $\Sp_{2n}\lr{\Z}$ is the image of the mapping class group ${\rm Mod}\lr{S_n}$ acting on $H_1(S_n,\Z)$, see for instance refs.~\cite{omeara_symplectic_1978} and~\cite[Prop.~6.3,Thm.~6.4]{FarbMargalit+2012}. .
As homeomorphisms of the surface $S_n$, Dehn twists preserve the intersection numbers of loops, which is also reflected in the preservation of commutativity of the corresponding symplectic transvections
\begin{equation}
    t_{\bs{\gamma}}\lrq{t_{\bs{\alpha}}, t_{\bs{\beta}}}t^{-1}_{\bs{\gamma}}=\lrq{t_{t_{\bs{\gamma}}\lr{\bs{\alpha}}}, t_{t_{\bs{\gamma}}\lr{\bs{\beta}}}}.
\end{equation}

\begin{figure}
\centering
\includegraphics[width=.5\textwidth]{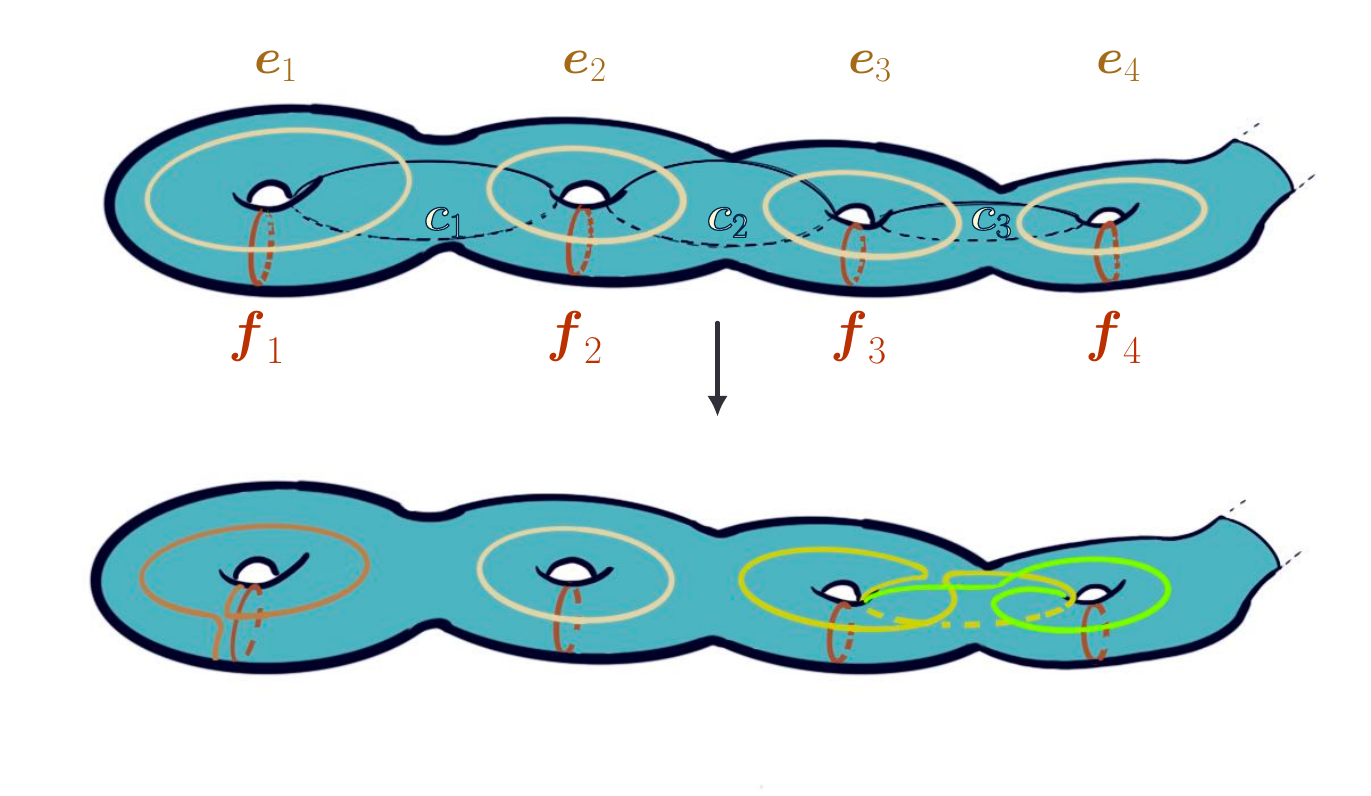}
\caption{A visualization of the surface $S_n$, where logical operators for the GKP code are represented as elements of the first homology group indicated by the elements $(\bs{e}_i, \bs{f}_i)$. 
Logical Clifford transformations are represented by sequences of Dehn twists of the torus in this representation, and we depict how two such transformations act on these generators going from the top to the bottom figures. 
On the left-most handle, we show how a Dehn twist about $\bs{f}_1$, a.k.a.\ a symplectic transvection $t_{\sqrt{d}\bs{f}_1}$, implements a logical phase gate by mapping $\bs{e}_1 \mapsto \bs{e}_1-\bs{f}_1$.
On the right-most handle, a logical $CZ$ gate (up to single-mode phase gates) is realized via a Dehn twist about the loop with label $\bs{c}_3$, which corresponds to a symplectic transvection $t_{\sqrt{d}(\bs{f}_3+\bs{f}_4)}$.
} \label{fig:Dehn}
\end{figure}

In fig.~\ref{fig:Dehn} we depict such a generating set (known as the \textit{Lickorish} generators \cite[Thm.~4.13]{FarbMargalit+2012}), where each Dehn twist label is associated with a lattice vector given either by a canonical basis element $e_i, f_i$ or a linear combination of these.

\medskip
\paragraph*{Example: the square lattice, $\CL=\sqrt{2}\Z^2$.}
For the single-mode square GKP code we can choose bases such that $M=2M^{\perp}=\sqrt{2}I$. 
Relative to this choice the first row of $M^{\perp}$ represents the logical $X$-type Pauli operator 
$\hat{X}=e^{-i\sqrt{\pi}\hat{p}}$ while the second represents $\hat{Z}=e^{i\sqrt{\pi}\hat{q}}$. 

Via eq.~\eqref{eq:AutS} we can identify a symplectic transformation $g$ that implements a non-trivial Clifford gate with its integral representation via  $U=g^T$.

It is convenient to introduce the $S$ and $T$ matrices,
\begin{equation}
    S\coloneqq\begin{pmatrix}
        0 & -1 \\ 1 & 0
    \end{pmatrix},\;\;
    T\coloneqq\begin{pmatrix}
        1 & 1 \\ 0 & 1
    \end{pmatrix},\label{eq:ST}
\end{equation}
which generate $\Sp_2\lr{\Z}=\langle S, T \rangle$.
In the integral representation, the $S$-matrix just introduced can be seen to implement a logical Hadamard gate $U_H=S$, while the logical phase gate $\hat{P}$ can be obtained from $U_P:= T$.
The $S$-matrix is orthogonal, just as the associated symplectic transformation on the lattice and thus the logical Hadamard can be implemented by a mere passive linear optical element with a representative Gaussian unitary $\hat{U}_H=e^{\pm i\frac{\pi}{2}\hat{n}}$ corresponding to a $\pi/2$ rotation in phase space. 
For the square lattice, the orthogonal symplectic automorphisms reduce modulo $2$ only to the identity and Hadamard actions. The phase action represented by $T$ therefore has no orthogonal implementation.

\medskip
\paragraph*{Example: the hexagonal lattice, $\CL=\sqrt{2}A_2$.}
For the hexagonal GKP code we have $M^{\perp}=M_{A_2}/\sqrt{2}$. 
As a root lattice, orthogonal automorphisms are given by reflections 
\begin{equation}
\label{eq:reflection}
    r_{\bs{\alpha}}\coloneqq I-2\frac{\bs{\alpha}\bs{\alpha}^T}{\bs{\alpha}^T\bs{\alpha}}
\end{equation}
along the roots $\bs{\alpha}$ and $\bs{\beta}$ contained in the rows of $M_{A_2}$ in eq.~\eqref{eq:MA2}.
Reflections are involutions with a $-1$ determinant, and hence are not symplectic. 
The symplectic orthogonal automorphisms thus form the rotation subgroup $C_6\cong\Aut^{SO}(A_2)$ of the full point group $D_6$, which contains the even subgroup of the Weyl group $W\lr{A_2}$ generated by the product of the two reflections
\begin{equation}
    R_{\frac{2\pi}{3}}=r_{\bs{\beta}}r_{\bs{\alpha}}=
    \begin{pmatrix}
        \cos \frac{2\pi}{3} & -\sin \frac{2\pi}{3} \\ 
        \sin \frac{2\pi}{3}& \cos \frac{2\pi}{3}
    \end{pmatrix}.
\end{equation}
Solving $UM_{A_2}=M_{A_2}R_{\frac{2\pi}{3}}^T$ yields the integral representation
\begin{equation}
U=\begin{pmatrix}
-1 & 1 \\ -1 & 0
\end{pmatrix}.\label{eq:Umat}
\end{equation}
By probing its effect on the standard basis, we find that this matrix implements the transformation $X\mapsto Y\mapsto Z\mapsto X$ on logical Pauli operators modulo phase. This is the projective Pauli action of the logical $\lr{\hat P\hat H}^{\dagger}$ gate~\cite{GCB}.

\section{GKP codes from compact Riemann surfaces}\label{sec:curves}

The connection between the symplectic automorphism group of a $2n$-dimensional symplectic lattice and the homological representation of the mapping class group of a compact genus-$n$ surface $S_n$ hints at a deeper relation between symplectic lattices and compact surfaces.
As we elaborate below, this connection motivates the study of GKP codes whose lattices arise as \emph{Jacobians} of algebraic curves. 
We begin by first clarifying the relation between logical operators of GKP codes and the homology of compact surfaces. 

The identification of symplectic lattice automorphisms with homology representations of transformations generated by Dehn twists, which are \textit{intersection number preserving} homeomorphisms of genus $n$ surfaces, suggests a more intuitive understanding of the topological nature of scaled GKP codes. 
An equivalent way to understand this connection is to choose symplectic bases giving an isomorphism of symplectic modules
\begin{equation}
    H_1\lr{\R^{2n}/\CL^{\perp}, \Z} \sim H_1\lr{S_n, \Z} \label{eq:Homology}
\end{equation}
under which $d$ times the physical symplectic inner product agrees with the oriented algebraic intersection pairing on $H_1\lr{S_n,\Z}$. 
Since we have $\CL=d\CL^{\perp}$, the torus $\R^{2n}/\CL$ is a $d^{2n}$-fold cover of $\R^{2n}/\CL^{\perp}$; treating cycles from $\CL$ as logically trivial reduces their coefficients modulo $d$. 
By eq.~\eqref{eq:Homology} we can thus regard elements in $H_1\lr{S_n, \Z_d} $ as representations of logical operators for the scaled code of type $dI_n$. 
A $d$-fold winding represents a stabilizer element, and the intersection number of two oriented cycles $[\gamma]$ and $[\eta]$ determines the phase $\exp\!\lr{-2\pi i([\gamma]\cdot[\eta])/d}$ of the ordered commutator $D_\gamma D_\eta D_\gamma^{-1}D_\eta^{-1}$

\subsection{Jacobians and compact Riemann surfaces}
The Jacobian can be thought of as a first-order approximation of a compact Riemann surface, which contains the information about the first homology group of the surface and the intersection between its elements.
We now outline the essential steps of the construction of the Jacobian and its associated symplectic lattice from a compact Riemann surface. For more detailed treatments see \cite[Ch.~11]{Birkenhake_2004} and \cite{Sarnak1994, Berge}.

Let $C$ have genus $n=\dim_{\C}H^0(\omega_C)$, where $H^0(\omega_C)$ denotes the space of holomorphic $1$-forms, and consider an oriented basis $\langle\gamma_1,\ldots,\gamma_{2n}\rangle_{\Z}=H_1(C,\Z)$ of closed loops with $\gamma_i\cdot\gamma_j=J_{ij}$.
Choosing a basis $\omega_1,\hdots, \omega_n$ for $H^0\lr{\omega_C}$ yields a linear map
\begin{equation}
    p: H_1\lr{C, \Z} \rightarrow \C^n :\; \gamma \mapsto \lr{\int_{\gamma} \omega_1, \hdots , \int_{\gamma} \omega_n}^T \label{eq:periods}
\end{equation}
defined from the set of $2n$ generators of $H_1\lr{C, \Z}$ to $2n$ vectors in $\C^n$. 
The $n \times 2n$ matrix 
\begin{equation}
    \Pi\coloneqq\begin{pmatrix} 
            p(\gamma_1)\; \hdots\; p(\gamma_{2n}) \label{eq:Jacobian}
        \end{pmatrix}
 \end{equation}
is known as the \emph{period matrix}. 
The period matrix admits a standard form. 
In particular, we can always choose a basis and normalization such that $\Pi$ takes the canonical form $\Pi=\lr{ I_n\; \Omega}$ where $\Omega$ is symmetric and $\Im \Omega > 0$ \cite[Sec.~8.1]{Birkenhake_2004}. 

Consider the lattice $\Lambda$ spanned by the columns of $\Pi$, $\Lambda\coloneqq\Pi\,\Z^{2n}$. 
The complex torus $T_{\Lambda} = \C^n/\Lambda$ obtained from the quotient by $\Lambda$ is known as the \emph{Jacobian} variety $J(C)$ of $C$. 
We can map $\Lambda$ into real space lattice $L_{\Lambda}$ by associating with each vector $\Lambda\ni\bs{v} \mapsto (\Re \bs{v}^T, \; \Im\bs{v}^T)^T$; this is known as the real representation. 
Writing $\Omega=X+iY$, $L_{\Lambda}$ is generated by the rows of the matrix $M$ defined by
\begin{equation}
M^T\coloneqq\begin{pmatrix}
I_n & X \\ 0 & Y
\end{pmatrix},
\end{equation} and satisfies
\begin{equation}
    M \lr{J_2 \otimes Y^{-1}} M^{T}=J.\label{eq:Msymp}
\end{equation}
Since $Y>0$, we can define the rescaled generator matrix
\begin{equation}
    M_C\coloneqq M(I_2\otimes Y^{-\frac{1}{2}}), \label{eq:MC}
\end{equation}
which, by eq.~\eqref{eq:Msymp}, is symplectic and generates a symplectic lattice, so that it can be scaled to yield a GKP code as discussed earlier.

Eq.~\eqref{eq:MC} also allows for an interpretation of the lattice generated by $M$, it is simply a stretched version of the symplectic lattice spanned by $M_C$ with ``stretching" $Y^{\frac{1}{2}}\oplus Y^{\frac{1}{2}}$. In the simple case where $Y=dI_n$, this becomes equivalent to the lattice present in a scaled GKP code of type $D=dI_n$. %
To obtain an alternative interpretation of the lattice $L_{\Lambda}$, we apply a generalized squeezing operation via $M_C\mapsto M_C'\coloneqq M_C\lr{Y^{-\frac{1}{2}} \oplus Y^{\frac{1}{2}}}$, which is again symplectic since $Y^{-\frac{1}{2}} \oplus Y^{\frac{1}{2}}\in\Sp_{2n}(\R)$ for symmetric $Y>0$,
and take the transpose to express
\begin{equation}
    M^T=(Y\oplus I_n) M_C'^T\,.
\end{equation}
Comparing this to eq.~\eqref{eq:scaledGKP} yields the interpretation of the lattice $L_{\Lambda}^T$ spanned by the generator matrix $M^T$ as a lattice corresponding to a GKP code of ``type'' $Y$, where we put type in quotes since there is no a priori reason for $Y$ to be integral or diagonal.

Let $\bs{r}_i$ denote the $i$th row of $\Pi$. Riemann's bilinear relations give
\begin{equation}
    \int_C\omega_i\wedge\overline{\omega}_j
    =\bs{r}_iJ\overline{\bs{r}_j}^{T}
    =\lr{\Pi J\overline{\Pi}^{T}}_{ij},
\end{equation}
and $i\Pi J\overline{\Pi}^{T}$ is positive definite; see Ref.~\cite[Ch.~4]{Birkenhake_2004}.
For $\Pi=(I_n\;\Omega)$ with $\Omega=X+iY\in\mathbb H_n$, the torus $\C^n/(\Z^n+\Omega\Z^n)$ is principally polarized by $H(x,y)=x^\dagger Y^{-1}y$, whose imaginary part is integral on the period lattice. ``Principal" means that the alternating form has invariant factors $(1,\ldots,1)$. %

Rather than to refer to compact Riemann surfaces, one typically refers to the Jacobian associated to a projective complex algebraic \emph{curve}. 
This underlies a deep connection between algebra and geometry: projective complex algebraic curves can be understood as ``explicit parametrizations" of compact Riemann surfaces. 
We now briefly outline this connection and refer the reader for a more detailed treatment to ref.~\cite{Griffiths1989, Bobenko2011}.

A complex algebraic curve $C=\lrc{(x,y)\in \C^2,\; P(x,y)=0 }\subset \C^2$ is the set of roots of an irreducible polynomial $P$ in two variables of total degree $d$. Its projective closure is $C_h=\lrc{[x:y:z]\in\CPP^2:\;P_h(x,y,z)=0}$, where $P_h(x,y,z)=z^dP(x/z,y/z)$ is the homogenization of $P$ and $[x:y:z]$ denotes homogeneous coordinates. There are finitely many singular points $S=\{(x_0, y_0)\in C:\; \partial_{x}P(x_0, y_0)=\partial_{y}P(x_0, y_0)=0 \}$. Away from them, the implicit-function theorem gives either $y=y(x)$, with $dy/dx=-(\partial_xP)/(\partial_yP)$ when $\partial_yP\neq0$ and $x$ a local coordinate, or $x=x(y)$, with $dx/dy=-(\partial_yP)/(\partial_xP)$ when $\partial_xP\neq0$ and $y$ a local coordinate.
By normalization and compactification, the nonsingular part $C^*=C\setminus S$ extends to a compact Riemann surface $\widehat C$\cite{Griffiths1989}. %
Conversely, Riemann showed that all compact Riemann surfaces can be described as compactifications of algebraic curves.

While these arguments show the one-to-one correspondence between compact Riemann surfaces and nonsingular projective algebraic curves, they are unwieldy in the explicit computation of the period integrals to construct Jacobians associated to curves. In the special case of hyperelliptic curves given by polynomials of the form
\begin{equation}
    P(x, y)=y^2-f(x),\hspace{3cm} f(x)=\prod_{i=1}^N (x-\lambda_i),\; \lambda_i\neq \lambda_j\; \forall\, i\neq j
\end{equation}
the identification of the homology basis of the corresponding compact Riemann surface (of genus $n=\lfloor (N-1)/2\rfloor$) and construction of the holomorphic forms become simpler:
Solutions to the curve are of the form $y=\sqrt{f(x)}$
where a homology basis with $2n$ generators is derived from the branch cuts of the complex square root spanned between the roots $\lambda_i$ (see \cite[p. 160]{Silverman2009} for an illustration). 
A basis for the holomorphic differentials is then given by
$\omega_i=x^{i-1}\frac{dx}{\sqrt{f(x)}}$ for $i=1\hdots n$ \cite{Birkenhake_2004}.

These connections allow us to associate (scaled) GKP codes with complex curves via their Jacobians. 
The chain of maps from curves to GKP codes via the Jacobian is pictured in figure \ref{fig:curvesGKP}.

\begin{figure}
\centering
\begin{tikzpicture}
\node (RC) at (0,0) {\rm Compact Riemann surface $\widehat{C}$};
\node (C) at (10,0) {\rm Curve $C$};
\node (JC) at (0,-2.5) {\rm Jacobian $J(C)$};
\node (L) at (5,-2.5) {\rm symplectic lattice $L(M_C)$};
\node (G) at (10,-2.5){\rm scaled GKP code $\CL_{\Lambda}$};
\draw[<-] (2.5,0.05) --node[above]{normalization theorem \cite{Griffiths1989}} (9,0.05);
\draw[->] (2.5,-0.05) --node[below]{Riemann existence theorem \cite{Griffiths1989}} (9,-0.05);
\draw[->] (JC) -- node[above] {eq. \eqref{eq:MC}} (L);
\draw[->]  (RC) -- node[right] {period mapping eqs. \eqref{eq:periods}, \eqref{eq:Jacobian} } (JC);
\draw[<->] (L) -- node[above]{eq. \eqref{eq:scaledGKP}}  (G);
\end{tikzpicture}
\caption{The chain of correspondences and maps that associate a GKP code to an irreducible complex algebraic curve $C\subset \CPP^2$.}\label{fig:curvesGKP}
\end{figure}
 
While these relationships have some interesting implications for GKP codes, the most immediate question is whether every GKP code can be understood as a curve. 
The answer to this question is known to be negative: there are symplectic lattices, such as the $E_8$ lattice, that do not arise as the Jacobian of curves \cite{Sarnak1994, Berge}. 
The general question of ``which principally polarized abelian varieties arise as Jacobians of curves" is a long-standing mathematical quest known as the \emph{Schottky problem} \cite{Sarnak1994, grushevsky2010schottky}. Despite their nice conception, codes in the Jacobian locus are constrained quantitatively: by the theorem of Buser and Sarnak~\cite{Sarnak1994}, the period lattice of a genus-$n$ surface satisfies $\lambda_1^2\leq\frac{3}{\pi}\log(4n+3)$, so that scaled GKP codes arising from Jacobians have $\Delta_{\rm GKP}\leq d^{-1/2}\sqrt{(3/\pi)\log(4n+3)}$, logarithmic in $n$, whereas general symplectic lattices reach $\lambda_1^2\sim n/(\pi e)$, which (up to constants) is the scaling necessary for families of ``good'' GKP codes~\cite{HarringtonPreskill, conrad2023good}.

\subsection{Single-mode GKP codes from elliptic curves}

In the previous section we have established a connection between complex curves and GKP codes which we now make more concrete for the case of a single-mode $n=1$. 
An \emph{elliptic curve} over $\C$ is a pair of a complex torus and a chosen origin $z\in\C/\Lambda$~\cite[Def.~1.1]{hain2014lectures}, 
\begin{equation}
    E=(\C/\Lambda, z),
\end{equation}
where $\Lambda \subset \C$ is a full-real-rank lattice. All such choices of origin are isomorphic via translation. In the context of GKP codes, phase space supplies a reference origin $0\in\C$, and the choice of origin $z$ relative to it selects the syndrome sector, the translation isomorphism being the displacement by $z$ that maps this sector onto the code space. In this sense $z$ encodes syndrome and logical-displacement data.

This definition of an elliptic curve is in bijection with an algebraic definition in the following sense. 
The curve $C_{g_2(\Lambda), g_3(\Lambda)}=\lrc{(x, y) \in \C^2, \; y^2=4x^3-g_2(\Lambda)x-g_3(\Lambda) }$, specified by two complex numbers $g_2, g_3$ that are the image of a lattice under the functions defined below, is parameterized by the Weierstrass $\wp$ function
\begin{equation}
    \wp\lr{z,\, \Lambda}:=\frac{1}{z^2}+\sum_{\omega\in \Lambda\setminus\lrc{0}}\lr{\frac{1}{(z-\omega)^2}-\frac{1}{\omega^2}},
\end{equation}
where the invariance under translations by lattice vectors $\wp(z+\Lambda)=\wp(z)$ shows that this is a well-defined function on the complex torus $\C / \Lambda$ with poles of order $2$ on each lattice point. 
Therefore, distinct lattices $\Lambda$, $\Lambda'$ are distinguished by their $\wp$ functions. 

The Weierstrass function $\wp$ also provides an alternative parametrization of the elliptic curve $C_{g_2(\Lambda),g_3(\Lambda)}$, which can be seen as follows. 
Introduce the (unnormalized) Eisenstein series of weight $k$
\begin{equation}
    G_k\lr{\Lambda}\coloneqq\sum_{\omega \in \Lambda\setminus \lrc{0}} \omega^{-k},
\end{equation}
and
\begin{equation}
    g_2(\Lambda)\coloneqq 60G_{4}\lr{\Lambda},\; g_3(\Lambda)\coloneqq 140 G_{6}\lr{\Lambda}\,.
\end{equation}
Then the equation for the elliptic curve is given by~\cite[Thm.~VI.3.5]{Silverman2009} 
\begin{equation}
    \wp'^{2}=4\wp^3-g_2\wp -g_3.
    \label{eq:elliptic_curve}
\end{equation}
The elliptic curve is non-singular, i.e.\ it has no cusps or self-intersections, when the discriminant of the right-hand side  
\begin{equation}
    \Delta(\Lambda) \coloneqq g_2^3-27g_3^2 
\end{equation}
is nonzero, which holds whenever $\Lambda$ is full-rank in $\C$. 

Let $\omega_1, \omega_2$ form a basis for the lattice $\Lambda=\omega_1 \Z \oplus \omega_2\Z$, which is full-rank if $\Im\lr{\omega_2/\omega_1}\neq 0$. 
One can fix an orientation of the basis elements by choosing a basis with $\Im\lr{\omega_2/\omega_1} > 0$, corresponding to a positive intersection of the homology element $\omega_1$ with $\omega_2$ on the torus $\C /\Lambda$, so that, up to an overall factor of rescaling and rotation $\omega_1$, the lattice  $\Lambda_{\tau}\coloneqq \Z  \oplus \tau \Z$ is parameterized by $\tau \in \hh:=\lrc{ z\in\C,\, \Im(z)>0}$ in the complex upper half plane. 
The Eisenstein series $G_k(\tau)$ has modular weight $k$~\cite[Sec.~2.1]{Zagier2008}; in particular, $g_2$ and $g_3$ above have weights $4$ and $6$. A modular form $f$ of weight $k$ obeys $f(\gamma.\tau)=(c\tau+d)^k f(\tau)$ for $\gamma\in\SL_2(\Z)$, where we have introduced the Möbius transformation
\begin{equation}
    {\begin{pmatrix}
        a & b \\ c & d
    \end{pmatrix}
    }.\tau \coloneqq \frac{a\tau+b}{c\tau+d}\,.
\end{equation}
As a function of $\tau$, the discriminant modular form $\Delta(\tau)\coloneqq\Delta(\Lambda_{\tau})$ is a modular cusp form of weight $12$, meaning in particular that it vanishes at the cusp $i\infty$~\cite[Sec.~2.4]{Zagier2008}. 
Möbius transformations with elements $\gamma=\begin{pmatrix}
a & b \\ c & d
\end{pmatrix}\in \SL_2\lr{\Z}$ can be understood as basis transformation of the corresponding lattice via the map
\begin{equation}
\begin{pmatrix}
1 \\ \tau 
\end{pmatrix}
\mapsto
X\gamma X
\begin{pmatrix}
1 \\ \tau 
\end{pmatrix}
=
(c\tau+d)\begin{pmatrix}
1 \\ \gamma.\tau 
\end{pmatrix},
\end{equation}
where $X=\begin{pmatrix}
0 & 1 \\ 1 & 0
\end{pmatrix}$ and $X\gamma X \in \SL_2\lr{\Z}$. 
For a fixed volume $\det \Lambda_{\tau}$, the lattice $\Lambda$ can always be recovered via appropriate rescaling up to a global rotation.

To associate a single-mode GKP code to an elliptic curve, recall that $\Sp_2(\R)=\SL_2(\R)$ acts transitively on the upper half plane $\hh$ by Möbius transformations. 
For symplectic orthogonal matrices $K\in \SO_2\lr{\R}=\Sp_2\lr{\R}\cap O_2\lr{\R}$, we have that $i$ is a fixed point, $i=K.i$. 
Therefore every point in the upper half plane $\tau=S.i\in \hh$ is in one-to-one correspondence with a coset $S\in \Sp_2\lr{\R}/\SO_2\lr{\R}$. 
We associate with the symplectically self-dual lattice $\Z^2$ the square GKP code encoding a qudit with dimension $d$ by rescaling the lattice $\Lambda_{\tau} \mapsto \sqrt{d/\det\lr{\Lambda_{\tau}}} \Lambda_{\tau}$.
Since this rescaling can always be done, it suffices to identify  $\Lambda_{\tau}$, equivalently the torus $\C/\Lambda_{\tau}$, with the corresponding qudit GKP code. 

The orbit $\Sp_2(\R).i$ gives all single-mode fixed-covolume lattice shapes modulo rotation. These shapes determine elliptic curves $E=\C/\Lambda_\tau$ while the marked point $z\in E$ separately records syndrome data.
First we interpret $\Lambda_{\tau}$ as the lattice associated to the stabilizer group of a GKP code. 
Then $z$, which labels a point in $\C$ up to a displacement by a stabilizer element in $\Lambda_{\tau}$, is interpreted as the sum of a syndrome $z\mod \frac{1}{d}\Lambda_{\tau}$ and a representative logical displacement label $\overline{z}\in \frac{1}{d}\Lambda_{\tau}$, where $\overline z$ denotes a chosen representative modulo $\Lambda_\tau$. See the discussion in Sec.~\ref{sec:Cliffords}. 

A level-$d$ structure~\cite[Def.~4.6]{hain2014lectures} is a symplectic basis $\lr{\frac{1}{d},\frac{\tau}{d}}$ of $E[d]\cong H_1(E,\Z_d)$. Its intersection pairing determines the Heisenberg--Weyl phase in Eq.~\eqref{eq:HW_operators}. Thus it records logical Pauli labels, while $z$ records syndrome and displacement data.

\begin{figure}
    \centering
    \includegraphics[width=.4\textwidth]{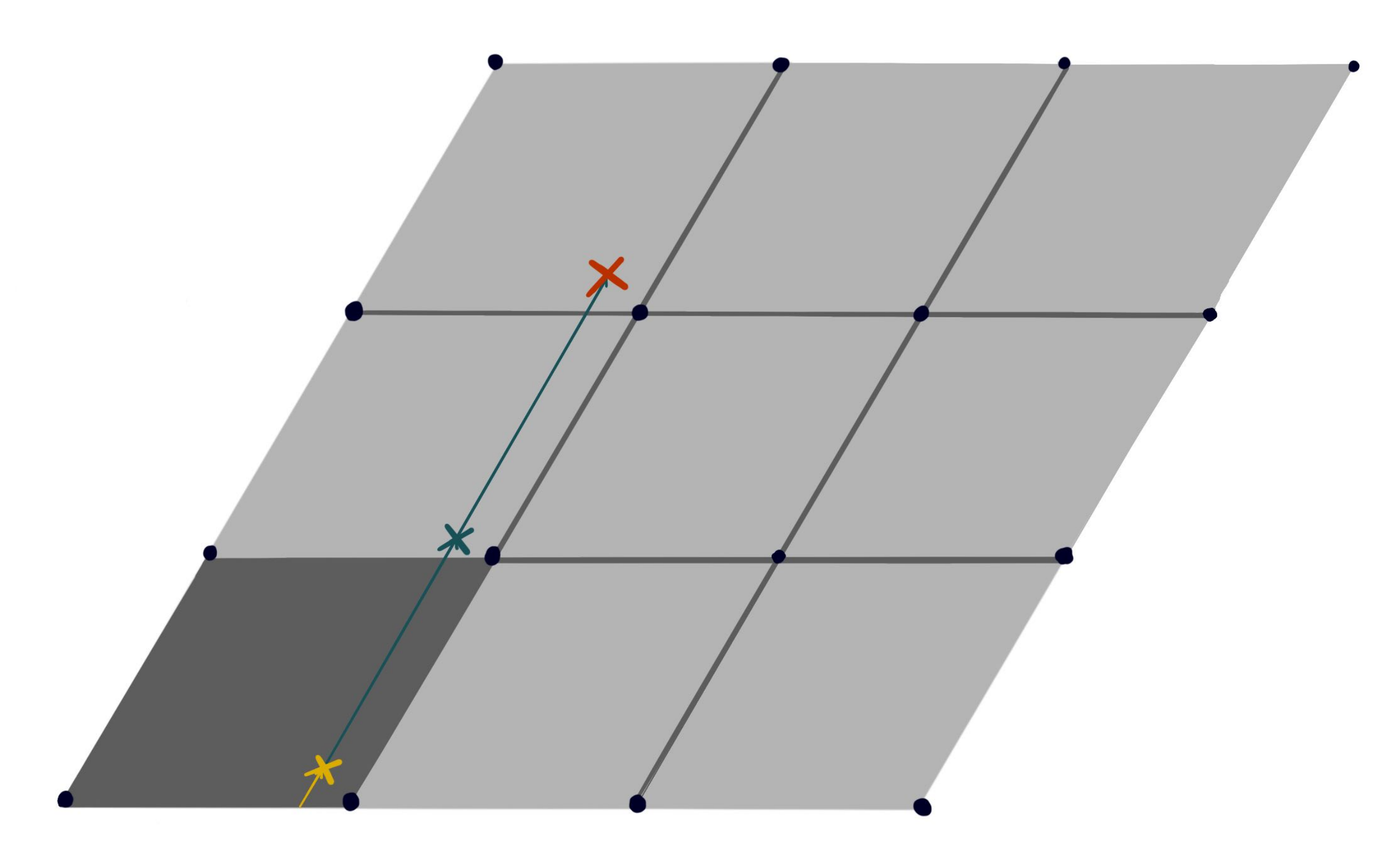}
    \caption{The hexagonal GKP code given by $\Lambda_{\rho},\; \rho=e^{i\pi/3}$ and relative level structure $d^{-1}\Lambda_{\rho}$ for $d=3$. The points $z\mod d^{-1}\Lambda_{\rho}$ parametrize the syndrome of the GKP code while the $d$-torsion points $d^{-1}\Lambda_{\rho}/\Lambda_{\rho}$ in $\C/\Lambda_{\rho}$ are interpreted to label logical Pauli operators for the associated GKP code.}
    \label{fig:torsion}
\end{figure}

\section{Moduli space of GKP codes and fiber bundle fault tolerance}\label{sec:GKP_FT}
\subsection{GKP distance and modular discriminant}

In this section, we compare uniform positive lower bounds on the distance of single-mode lattice realizations with lower bounds on the modular discriminant.

\begin{figure}
\includegraphics[width=.6\columnwidth]{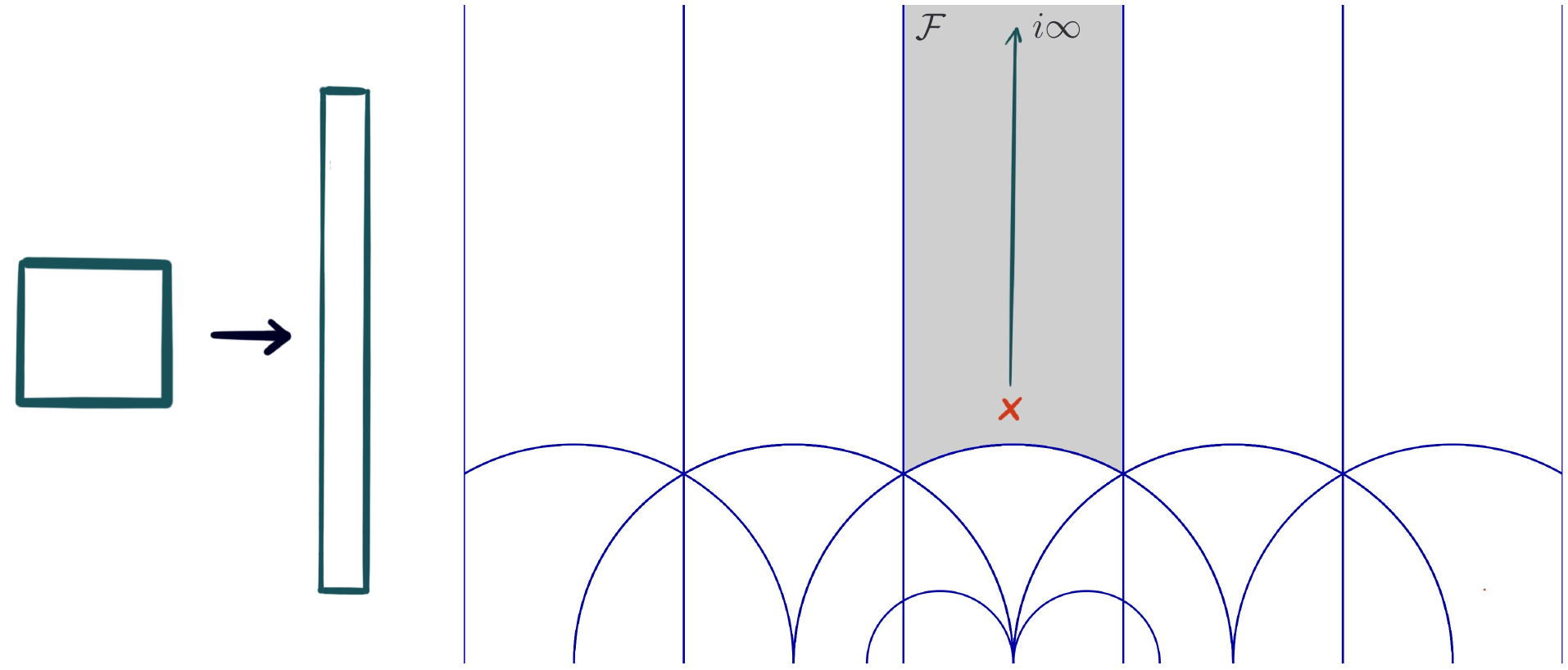}
\caption{The fundamental domain $\CF$ is marked in grey on the RHS. 
We illustrate the effect of a squeezing operation $\tau \mapsto (\lambda \oplus \lambda^{-1}).\tau=\lambda^2 \tau$ and the corresponding transformation on the lattice $\Lambda_{\tau} \mapsto \Lambda_{\lambda^2\tau}/ \sqrt{\det\lr{\Lambda_{\lambda^2\tau}}}$.}\label{fig:squeeze_tess}
\end{figure}

To distinguish lattice realizations from lattice shapes, the space of oriented fixed-covolume lattice realizations modulo integral basis change is $\Sp_2(\Z)\backslash\Sp_2(\R)$. Quotienting additionally by global rotations gives
\begin{equation}
M_1\coloneqq\PSp_2(\Z)\backslash\mathbb H
\cong\Sp_2(\Z)\backslash\Sp_2(\R)/\SO_2(\R),
\end{equation}
which parametrizes lattice shapes, equivalently isomorphism classes of complex elliptic curves~\cite[Prop.~1.17 and Def.~3.12]{hain2014lectures}. The coarse analytic space underlying this moduli orbifold is a Riemann surface, with modular $j$-invariant
\begin{equation}
j(\tau)\coloneqq 1728\frac{g_2^3(\tau)}{\Delta(\tau)}.\label{eq:j_def}
\end{equation} 
We can represent $M_1$ via the fundamental domain
\begin{equation}
\CF\coloneqq\lrc{\tau \in \hh: \; |\Re\lr{\tau}|\leq \frac{1}{2},\; |\tau|\geq 1},
\end{equation}
shown in fig.~\ref{fig:squeeze_tess}.

The moduli space $M_1$ constructed should be understood as an orbifold as the $\PSp_2(\Z)$ action is not free. With $S.z=-1/z$ and $T.z=z+1$, the points $i$ and $\rho:=e^{i\pi/3}$ are fixed by $S$ and $ST^{-1}$, respectively. These correspond to the square and hexagonal lattices and admit particular passive logical Cliffords. 
We begin by investigating the connection between the topology of $M_1$ and the coding theoretic properties of the associated codes.

The $j$ function diverges in the limit $\tau\rightarrow i\infty$. 
Using $q=e^{i2\pi \tau}$, we can write
\begin{align}
j\lr{\tau}&=q^{-1}+744+196884q + \hdots,\\
\Delta\lr{\tau}&=(2\pi)^{12}q\prod_{k=1}^{\infty}(1-q^k)^{24}=(2\pi)^{12}\sum_{n=1}^{\infty} \tau(n)q^n,\\
g_2\lr{\tau}&=\frac{4\pi^4}{3}\lr{1+240\sum_{n=1}^{\infty}\sigma_3(n)q^n},
\end{align}
where we have the Ramanujan $\tau(n)$ function and the divisor sum $\sigma_3(n)=\sum_{d|n}d^3$ function~\cite[Sec.~2]{Zagier2008}. The source of this divergence is the simple root of $\Delta(\tau)$ in the limit $\tau\rightarrow i\infty$. 
Recall from the previous section that $\Delta(\Lambda)\neq0$ whenever $\Lambda$ is full rank, such that the divergence corresponds to the limit where the lattice $\Lambda_{\tau}$ is not full rank anymore. To understand this point better, write $\tau=x+iy=M.i$, with
\begin{equation}
M\coloneqq\begin{pmatrix}
\sqrt{y} & x/\sqrt{y} \\ 0 &1/\sqrt{y}
\end{pmatrix}.
\end{equation}
The shortest vector in the lattice $L(M)$ spanned by the rows of $M$ satisfies
\begin{align}
\lambda_1^2\lr{L(M)} &=\min_{(0,0)\neq(n, m) \in \Z^2} \left(n^2y+\frac{(nx+m)^2}{y}\right) \nonumber\\&\leq \min\left\{y+\frac{x^2}{y},\, \frac{1}{y}\right\},
\end{align}
so that in particular we have $\Im\lr{\tau}\leq\lambda_1^{-2}(L(M))$. 
That is, representing the lattice basis in $\hh$, demanding that the lattice (the corresponding GKP code) has finite non-zero distance $\lambda_1\geq const.$ yields an upper bound on the imaginary part of its representation in $\hh$. 
Similarly, the squeezing value ${\rm sq}\lr{M}:=\|M^T\|_2$ needed relative to the square lattice satisfies ${\rm sq}\lr{M}\geq \sqrt{\Im\lr{\tau}}$. 
We illustrate the intuition behind the limit $\tau\rightarrow i\infty$ being associated to a zero distance GKP code in fig.~\ref{fig:squeeze_tess}, where, starting at a square GKP code, a squeezing deformation maps $\tau=i\mapsto \lambda^2i,\, \lambda>0$. 
While one of the lattice basis vectors gets increasingly longer, due to the volume-preserving nature of $\Sp_2\lr{\R}$ the other shrinks until it converges to $0$ in the infinite squeezing limit. 

For $\tau=x+iy$, define the intrinsic Petersson norm of the modular discriminant by $\|\Delta\|(\tau):=y^6|\Delta(\tau)|$ and let $L_\tau=y^{-1/2}(\Z+\tau\Z)$ denote the associated covolume-one lattice.
\begin{prop}[Distance and discriminant]\label{prop:distance}
For every nonempty set $T\subseteq\hh$,
\begin{equation}
\inf_{\tau\in T}\lambda_1(L_\tau)>0
\quad\Longleftrightarrow\quad
\inf_{\tau\in T}\|\Delta\|(\tau)>0.
\end{equation}
For the associated scaled single-mode GKP codes with fixed integer $d\geq2$, these conditions are also equivalent to
\begin{equation}
\inf_{\tau\in T}\Delta_{\rm GKP}(\tau)>0.
\end{equation}
\end{prop}
\begin{proof}
Both $\|\Delta\|$ and $\lambda_1(L_\tau)$ are $\SL_2(\Z)$-invariant, so we may represent the points of $T$ in the standard fundamental domain $\CF$. There the shortest vector of $\Z+\tau\Z$ has length $1$, and hence $\lambda_1(L_\tau)=y^{-1/2}$. Thus $\lambda_1$ is bounded away from zero on $T$ exactly when $y$ is bounded above. On every truncated fundamental domain $\CF\cap\{y\leq Y\}$, the continuous and nowhere-vanishing function $\|\Delta\|$ has a positive minimum. Conversely, the product expansion gives $\|\Delta\|(\tau)\leq C y^6e^{-2\pi y}\to0$ uniformly as $y\to\infty$. This proves the equivalence. The final statement follows from $\Delta_{\rm GKP}=d^{-1/2}\lambda_1$ for scaled codes.
\end{proof}

\subsection{Topological interpretation: the trefoil defect}
We can understand the resulting space of GKP codes topologically via an interpretation presented in \cite{Ghys} and \cite{Milnor+1972}. 
After the weighted rescaling $g_2(c\Lambda)=c^{-4}g_2(\Lambda)$ and $g_3(c\Lambda)=c^{-6}g_3(\Lambda)$, the normalized parameters lie on $S^3$. The discriminant-zero degeneration $g_2^3-27g_3^2=0$ cuts out the trefoil knot $K$. Thus the space of oriented fixed-covolume single-mode lattice realizations modulo integral basis change is the complement $S^3-K\cong\Sp_2(\Z)\backslash\Sp_2(\R)$ for which the knot is an excluded zero-distance limit.
We illustrate the trefoil knot in fig.~\ref{fig:trefoil} and refer to ref.~\cite{AMSGhys, Ghys} for further reference. 

\begin{figure}
    \centering
    \includegraphics[width=.4\textwidth]{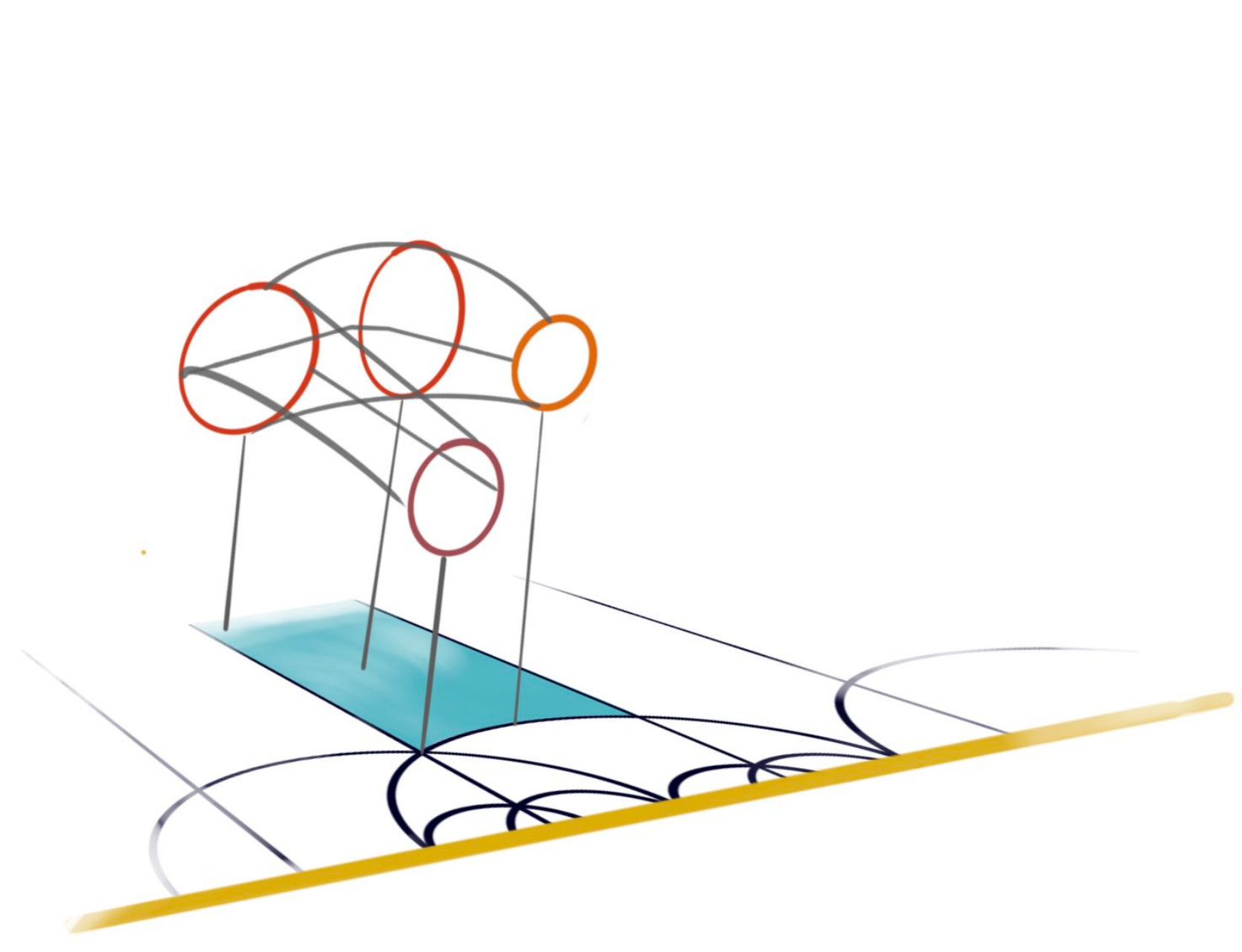}
    \caption{The Seifert fibration describing a decomposition of $\Sp_2\lr{\Z}\backslash \Sp_2\lr{\R}$ into the fundamental domain $\CF$ and a rotation label in $S^1$ for each point in $\CF$. While every lattice has a $\pi$-rotation symmetry, there are special (singular) points $i$ and $\rho=e^{i\pi/3}$ with additional symmetries under $\pi/2$  and $\pi/3$ rotation. This can be pictured by a smaller circumference rotation index attached to these points in the fibration. In terms of GKP codes,  it is these singular points on $\CF$ that correspond to GKP codes with additional orthogonal symplectic lattice automorphisms.}
    \label{fig:seifert}
\end{figure}

Any smooth implementation of a Clifford gate on a GKP code naturally traverses a continuous closed loop in the space of lattices $\Sp_2\lr{\Z}\backslash \Sp_2\lr{\R}$ while implementing a basis transformation. 
The topological defect in this space carved out by the trefoil knot illustrates that such loops are in general homotopically non-trivial.
One way to understand this is through the equivalence $\Sp_2\lr{\Z}\backslash \Sp_2\lr{\R}/\SO_2\lr{\R}=\Sp_2\lr{\Z}\backslash \hh$, represented by the fundamental domain $\CF$ (recall Sec.~\ref{sec:GKP_FT}). 
The space of lattices, up to a rotation, is labeled by an element in the fundamental domain so that each lattice -- including a rotation label --  can be labeled by a point in the fundamental domain $\CF$ \textit{together} with a rotation label in $S^1$ (which may vary across points in $\CF$). 
In order for a smooth transformation on the space of lattices to return to the same point in the fundamental domain with the same rotation label, it must either map to an $\SL_2\lr{\Z}=\Sp_2\lr{\Z}$ equivalent point in $\hh$, or perform a full rotation in $S^1$. 
This decomposition of $\Sp_2\lr{\Z}\backslash \Sp_2\lr{\R}$ is the so-called \textit{Seifert fibration} \cite{Seifert}, which we illustrate in fig.~\ref{fig:seifert}. 

In fact, the fundamental group of this space is the braid group on three strands, $\pi_1\lr{S^3-K}\cong B_3$ \cite{Gannon2023}. 
To see this, we return to label the lattice $\Lambda=\omega_1\Z +\omega_2\Z$ by the complex basis $\lr{\omega_1, \omega_2}$ for a minute. Since $\C$ is algebraically closed, the defining equation of the elliptic curve takes the form \cite[Sec.~III.1]{Silverman2009}
\begin{align}
    \wp'^2 &= 4\lr{\wp-e_1}\lr{\wp-e_2}\lr{\wp-e_3},\\
    \Delta&=16\lr{e_1-e_2}^2\lr{e_2-e_3}^2\lr{e_1-e_3}^2 \neq 0,
\end{align}
where $e_1=\wp\lr{\omega_1/2}, e_2=\wp\lr{\omega_2/2}$ and $e_3=\wp\lr{\lr{\omega_1+\omega_2}/2}$ with $e_1+e_2+e_3=0$ form the three distinct roots of the equation $\wp'=0$. Since $\wp\lr{z+\Lambda}=\wp\lr{z}$ is defined modulo the lattice and the coefficients $g_2, g_3$, as well as the lattice, are uniquely determined by the roots $e_1, e_2, e_3$, any smoothly parametrized basis transformation can also be identified by the evolution $t: [0,1]\rightarrow e_1(t), e_2(t), e_3(t)$, which smoothly implements a permutation of the three roots. Away from the trefoil defect $\Delta=0$, the positions of the variables determining roots on $\C/\Lambda$ remain distinct along the path, so that every non-trivial basis transformation implemented in this fashion can be identified with a non-trivial element in the braid group of three strands $B_3$ which has a representation in $\SL_2\lr{\Z}=\langle T, T^{-T} \rangle$ \cite{Gannon2023}, where $T$ is the shear of eq.~\eqref{eq:ST} and the two standard braid generators of $B_3$ map to these two shears. The following statement can be understood as a special case of a statement discussed in ref.~\cite{Burchards_GeometricGKP}.
\begin{prop}[Symplectic monodromy of lattice loops]\label{prop:noncontractible}
Let $G=\Sp_2(\R)$, $\Gamma=\Sp_2(\Z)$, $\mathcal X=\Gamma\backslash G\cong S^3-K$, and let
$p:G\to\mathcal X$ be the quotient map; fix $g_0\in G$ and set $x_0=p(g_0)$.
Then $p$ is a regular covering with deck group $\Gamma$. For a loop $\ell$ based at
$x_0$ define its monodromy $\mu_{g_0}(\ell)\in\Gamma$ by requiring that the lift of
$\ell$ starting at $g_0$ ends at $\mu_{g_0}(\ell)\,g_0$. Then $\mu_{g_0}(\ell)$ depends
only on the homotopy class of $\ell$, and $\mu_{g_0}$ is a surjective homomorphism
fitting into the exact sequence
\begin{equation}
1\longrightarrow\pi_1(G,g_0)\xrightarrow{\;p_*\;}\pi_1(\mathcal X,x_0)
\xrightarrow{\;\mu_{g_0}\;}\Gamma\longrightarrow1,\qquad \pi_1(G,g_0)\cong\Z.
\end{equation}
In particular, a loop with $\mu_{g_0}(\ell)\neq I$ is noncontractible; this applies to every loop with nontrivial symplectic monodromy modulo $d$, and hence to every Clifford implementation of a logically nontrivial Clifford gate.
\end{prop}
\begin{proof}
The discrete group $\Gamma$ acts freely and properly discontinuously on the connected Lie group $G$ by left translations, so $p$ is a regular covering with deck group $\Gamma$~\cite[Prop.~1.40]{Hatcher}. The lift of a loop $\ell$ based at $x_0$ that starts at $g_0$ ends in the fiber $p^{-1}(x_0)=\Gamma g_0$, and since $\Gamma$ acts freely this endpoint is $Ag_0$ for a unique $A=\mu_{g_0}(\ell)\in\Gamma$. By homotopy lifting, $\mu_{g_0}(\ell)$ depends only on the class of $\ell$. Since deck transformations map lifts to lifts, the lift of a concatenation $\ell_1\ell_2$ from $g_0$ is the lift of $\ell_1$ followed by $\mu_{g_0}(\ell_1)$ applied to the lift of $\ell_2$, so it ends at $\mu_{g_0}(\ell_1)\mu_{g_0}(\ell_2)\,g_0$ and $\mu_{g_0}$ is a homomorphism. It is surjective because $G$ is connected: a path from $g_0$ to $Ag_0$ projects to a loop with monodromy $A$. Its kernel consists precisely of the loops whose lifts close up, namely $p_*\pi_1(G,g_0)$, and $p_*$ is injective for a covering, which makes this sequence exact. Finally, the Iwasawa decomposition (Appendix~\ref{App:Moebius}) gives $\pi_1(G)\cong\pi_1(\SO_2(\R))\cong\Z$. A null-homotopic loop has trivial monodromy, which yields the last statement.
\end{proof}

\begin{figure}
    \centering
      \includegraphics[width=.5\textwidth]{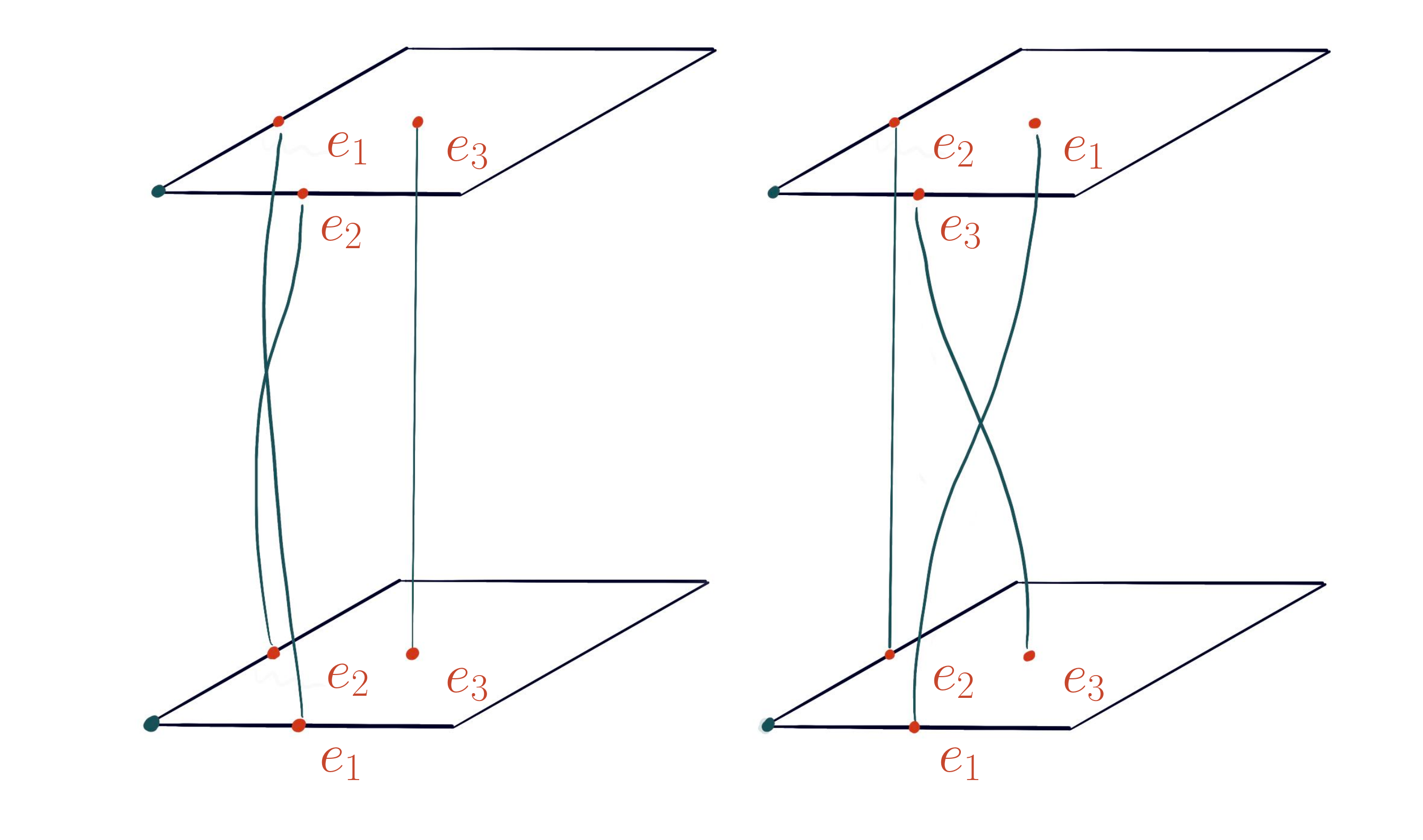}
    \caption{A braid on $e_1=\wp\lr{\omega_1/2}, e_2=\wp\lr{\omega_2/2}, e_3=\wp\lr{\lr{\omega_1+\omega_2}/2}$ implemented through a rotation $(\omega_1,  \omega_2)\mapsto (\omega_2, -\omega_1)$ (l.) corresponding to a GKP Hadamard gate and a shear $(\omega_1, \omega_2)\mapsto (\omega_1+\omega_2, \omega_2)$ (r.)  in the case of $d=2$. Nontrivial braiding of the three roots $e_i$ is induced by smoothly parametrized automorphisms of the underlying lattice.}
    \label{fig:braid}
\end{figure}

We illustrate the braids induced by a rotation and a shear on the square lattice -- corresponding to a logical Hadamard- and phase gate for the square GKP code with $d=2$ -- in fig.~\ref{fig:braid}.
In prop.~\ref{prop:noncontractible}, changing the reference lift from $g_0$ to $Bg_0$, with $B\in\Gamma$, conjugates the monodromy by $B$. Note, however, that non-contractibility of a loop does not imply non-trivial associated monodromy. For exmple, the half-rotation loop $t\mapsto p(g_0R_{\pi t})$ has monodromy $-I$, whose reduction modulo $2$ is trivial, while the full-rotation loop $t\mapsto p(g_0R_{2\pi t})$ has monodromy $I$ while both are noncontractible. These paths use right multiplication by $R_\phi$ and hence rotate the row lattice by $-\phi$.

In the following we introduce a linking number for loops through $\mathcal{X}$, which presents a homology- and homotopy invariant of the implementation loop.
The linking number is defined via a discriminant function $\widetilde{\Delta}:\Sp_2\lr{\Z}\backslash \Sp_2\lr{\R} \rightarrow \C^{\times}$ which provides an isomorphism of the homology groups $H_1\lr{\Sp_2\lr{\Z}\backslash \Sp_2\lr{\R}, \, \Z} \sim H_1\lr{\C^{\times}, \Z}$~\cite{Duke, Ueki}, so that closed loops in the space of symplectic lattices map to closed loops in $\C^{\times}$. 
The discriminant function $\widetilde{\Delta}$ is an invariant of the associated lattice, independent of the choice of basis, defined for $\gamma=\begin{pmatrix}
a & b \\ c & d
\end{pmatrix} \in \Sp_2\lr{\R}$
as
\begin{equation}
    \widetilde{\Delta}\lr{\gamma}\coloneqq j_{12}(\gamma,i)\Delta(\gamma.i),
\end{equation}
where we have defined the factor of automorphy $j_{12}(\gamma,z)\coloneqq(cz+d)^{-12}$.  
Now let $\gamma_A(t): [0,1]\rightarrow \Sp_2\lr{\R}$ be a continuous curve with $\gamma_A(0)\in \Sp_2\lr{\R}$ and $\gamma_A(1)=A\gamma_A(0)$, where $A\in \Sp_2\lr{\Z}$. The linking number is defined by
\begin{equation}
    \mathrm{link} \lr{\gamma_A, K}\coloneqq\frac{1}{2\pi i}\oint_{\gamma_A} \frac{\mathrm{d}\widetilde{\Delta}}{\widetilde{\Delta}}=\frac{1}{2\pi i}\oint_{\gamma_A} \frac{\mathrm{d}\Delta}{\Delta}+\frac{1}{2\pi i}\oint_{\gamma_A} \frac{\mathrm{d}j_{12}}{j_{12}}
    \label{eq:link}
\end{equation}
and is a topological invariant of the path \cite{Duke,Ueki}. Physically, it distinguishes homology classes of continuous implementations of a gate. Note that this winding number is not fixed by the monodromy representation. As in the example above, concatenation with the full-rotation loop $t\mapsto p(g_0R_{2\pi t})$ leaves the monodromy $A$ unchanged but changes the linking number by $-12$.
\medskip
\paragraph*{Example: the square GKP code, $\CL=\sqrt{2}\Z^2$.}
Recall from Sec.~\ref{sec:Cliffords} that the two basic symplectic automorphisms of the square GKP code are firstly the Hadamard gate with integral representation $S$ implemented by a $-\pi/2$ physical rotation and second the phase gate with integral representation $T$. 

A physical rotation $\Lambda\mapsto e^{i\phi}\Lambda$ is represented by $M\mapsto MR_\phi^T$, and the automorphy factor gives $\widetilde\Delta(MR_\phi^T)=e^{12i\phi}\widetilde\Delta(M)$. Whenever the rotation through $\pi/k$ closes the lattice path, its linking number is therefore $6/k$. In particular, for the Hadamard gate, a $\pi/2$ rotation of the square lattice has monodromy $S^{-1}$ and linking number ${\rm link}(\gamma_{S^{-1}},K)=3$. The reverse path has monodromy $S$ and linking number $-3$. See fig.~\ref{fig:trefoil} for an illustration of the positive rotation loop.

A continuous parametrization of a loop corresponding to the phase gate can be obtained by $t\in \lrq{0,1}:\;M\lr{t}=\begin{pmatrix}
    1 & t\\ 0 & 1
\end{pmatrix}M_0 \in \Sp_2\lr{\R}$,
where $M_0$ is the normalized generator matrix of the initial code. In the upper half plane, this path in $\Sp_2\lr{\R}$ is represented by $\tau\lr{t}=M\lr{t}.i=\tau_0+t$, where $\tau_0=M_0.i$ represents the initial lattice. Modulo the left $\Sp_2\lr{\Z}$ action, this path corresponds to a horizontal loop winding around the fundamental domain $\CF$ with linking number
\begin{equation}
    {\rm link}\lr{\gamma_T, K}=\frac{1}{2\pi i}\oint_{\gamma_T} \frac{\mathrm{d}\Delta}{\Delta}=\int_0^1 dt\, E_2\lr{\tau_0+t}=1
\end{equation}
where $E_2\lr{z}=1-24e^{i2\pi z} + \hdots$ is the second Eisenstein series normalized such that $E_2\lr{z \rightarrow i\infty} = 1$. The last equality follows from $\int_0^1 dt\, e^{i2\pi n t} =\delta_{0,n}$.
\medskip

\paragraph*{Example: the hexagonal GKP code, $\CL=\sqrt{2}A_2$.}
As for the square GKP code, the $2\pi/3$ rotation symmetry of the hexagonal lattice implements a Clifford gate with integral representation $U$, given in eq.~\eqref{eq:Umat}. Using the general argument presented above this corresponds to a linking number ${\rm link}\lr{\gamma_U,  K}=4$ of the corresponding path with the trefoil knot, while the $\pi/3$ rotation has linking number $2$.
\medskip

In a seminal paper, Ghys~\cite[Sec.~3.3]{Ghys} showed that for primitive hyperbolic elements $A\in\Sp_2(\Z)$ with $\operatorname{Tr}(A)>2$ which are implemented via a symplectic squeezing operation
\begin{equation}
    M\in \SL_2\lr{\R} \mapsto M (\lambda \oplus \lambda^{-1}) = AM,\; \lambda>1,
\end{equation}
the corresponding closed modular geodesic associated with the conjugacy class of $A$ has linking number ${\rm link}(\gamma_A,K)=\psi(A)$, where $\psi(A)$ is the Rademacher symbol. For the standard reduced positive alternating form $A\sim R^{r_1}L^{l_1}\cdots R^{r_N}L^{l_N}$, where $R=T$, $L=T^T$, and $r_i,l_i>0$, one has
\begin{equation}
\psi(A)=\sum_{i=1}^N(r_i-l_i).\label{eq:RL}
\end{equation}
The right-hand side is the difference of the exponents~\cite[Sec.~3.5]{Ghys}. The Rademacher symbol is a class invariant for $\Sp_2(\Z)$, which is further discussed in Appendix~\ref{app:Rademacher}. An implementation path may differ from $\gamma_A$ by additional loops, so its linking number is generally computed by Eq.~\eqref{eq:link}.

\subsection{The garden of GKP codes}
So far we have identified scaled single-mode GKP code shapes with elliptic curves with level-$d$ structure and identified the topological defect in the space of oriented fixed-covolume lattice realizations with the zero-distance limit. We have also shown how specified integral implementations of logical Clifford gates carry a linking number with this defect.

We now construct the finite quotient parametrizing single-mode scaled GKP code shapes with nonzero syndrome. This quotient can contain points with nontrivial finite symmetries and is therefore naturally an orbifold. Away from such points, the quotient map is an ordinary covering; Sec.~\ref{sec:fiber} discusses its monodromy.

Similarly to the case of generic elliptic curves (recall $M_1=\PSp_2(\Z)\backslash\hh$ from Sec.~\ref{sec:GKP_FT}), isomorphism classes of elliptic curves with level structure are classified by the quotient space $\mathcal{M}_1\lrq{d}=\Gamma(d)\backslash\hh$, where
\begin{equation}
\Gamma(d):=\ker\lr{\Sp_2(\Z)\longrightarrow\Sp_2(\Z_d)}
\end{equation}
is the principal congruence subgroup.

Define the action of $(m, n ) \in \Z^2$ on $\lr{\tau, z}$ as $\lr{\tau, z+m+n\tau} $, so that the quotient $ \Z^2\backslash\lr{\tau, z}$ is translation symmetric under translations of $z$ by elements in $\Lambda_{\tau}$ and define the $\Gamma\lr{d}$ action as 
\begin{equation}
\gamma:\; (\tau, z) \mapsto \lr{\gamma.\tau, z/(c\tau+d)}
\end{equation}
for $\gamma=\begin{pmatrix}
a & b \\ c & d
\end{pmatrix} \in \Gamma(d)$.
See~\cite[Sec.~2.3]{hain2014lectures} for background.

The action of elements in $\Gamma(d)$ preserves the level-$d$ structure $ d^{-1}\Lambda_{\tau}$ sitting inside of $\C/\Lambda_{\tau}$ and hence represents logically trivial basis transformations of the GKP code.
We assemble the orbifold family of elliptic curves with level-$d$ structure (single-mode GKP codes) as
\begin{equation}
    E\lr{d}\coloneqq\lr{\Gamma(d) \ltimes \Z^2}\backslash \hh \times \C.
\end{equation}
Understood as GKP codes, this is the scaled single-mode family of fixed type $d$; for each $\tau$, the point $z$ labels the syndrome and logical displacement.
For $d\geq3$, $\Gamma(d)$ is torsion-free and acts freely on $\hh$, so that $\Gamma(d)\backslash\hh$ is a fine moduli space. For $d=2$, however, the moduli space retains the stabilizer $\{\pm I\}$ because $-I$ acts trivially on the level-$2$ structure~\cite{hain2014lectures}.

\begin{figure}
\includegraphics[width=.6\columnwidth]{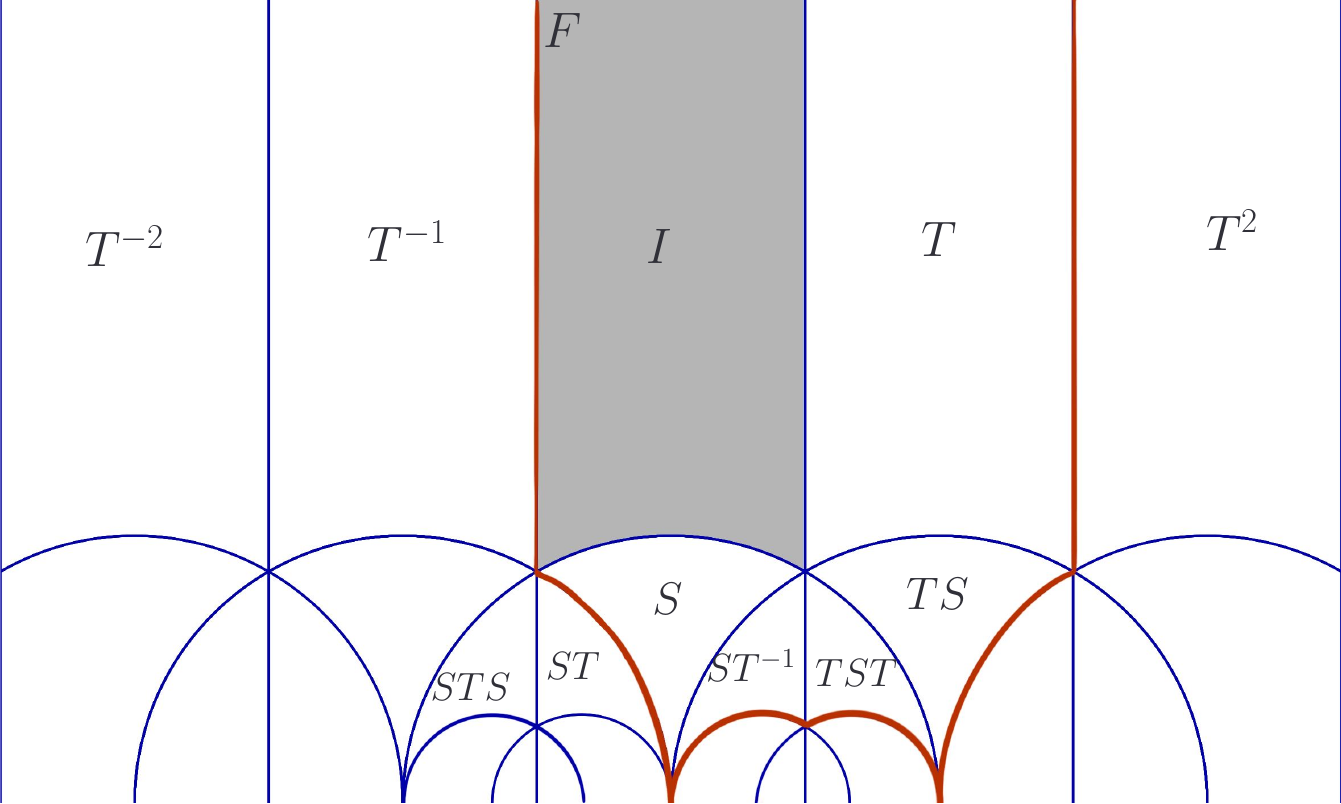}
\caption{The fundamental region $F(2)=\Gamma(2)\backslash\hh = \cup_{\gamma \in \Sp_2\lr{\Z_2}} \gamma F$ is drawn in red and contains the logical Clifford translates of the fundamental region $F=\SL_2\lr{\Z}\backslash \hh$. }\label{fig:F2}
\end{figure}

Let $\mathcal{T}_d\coloneqq\lrc{(\tau,z)\in\hh\times\C:\; z\in d^{-1}\Lambda_\tau}$ denote
the $d$-torsion locus; it is invariant under $\Gamma(d)\ltimes\Z^2$ and descends to
$E[d]\subset E(d)$. We define
\begin{equation}
E^{\times}\lr{d}\coloneqq\lr{\Gamma(d)\ltimes\Z^2}\backslash\lr{\hh\times\C\setminus\mathcal{T}_d},
\label{eq:Ex}
\end{equation}
For $d\geq3$, the fiber over $\tau$ is $E_\tau\setminus E_\tau[d]$, with zero syndrome removed. For $d=2$, its fiber is instead $(E_\tau\setminus E_\tau[2])/\{\pm1\}$. The family carries an action of the affine label group $G_d:=\Sp_2(\Z_d)\ltimes E[d]$, with $E[d]\cong(\Z/d\Z)^2$. We quotient by this affine label group to forget the level and displacement labels.

We define the quotient map for $d\geq 2$ by 
\begin{equation}
    E^{\times}(d) \xrightarrow{\pi} M^{\times}:=G_d\backslash E^{\times}(d)\,.
    \label{eq:EMspace}
\end{equation}
Here $E[d]$ acts by torsion translation and $\Sp_2(\Z_d)$ changes the level basis. Write $E^{\times}_{\rm free}(d):=\{x\in E^{\times}\lr{d}:\operatorname{Stab}_{G_d}(x)=1\}$ and $M^{\times}_{\rm free}:=G_d\backslash E^{\times}_{\rm free}(d)$. The following proposition follows from standard arguments for covering spaces.

\begin{figure}
\centering
\includegraphics{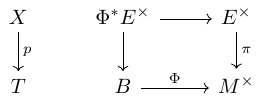}
\caption{After excluding points with nontrivial $G_d$-symmetry, the restricted quotient map $E^{\times}_{\rm free}(d)\to M^{\times}_{\rm free}$ is a $G_d$-covering. Any map $f:B\to M^{\times}_{\rm free}$ pulls this covering back to a $G_d$-covering over $B$.}
\label{fig:universal_family}
\end{figure}

\begin{prop}[Finite covering and symplectic monodromy]\label{prop:monodromy}
Let $d\geq 2$. The group $G_d=\Sp_2(\Z_d)\ltimes E[d]$ acts effectively on $E^{\times}(d)$,
where $E[d]$ acts by torsion translations and $\Sp_2(\Z_d)$ by changes of the level basis,
and the free locus $E^{\times}_{\rm free}(d)$ is connected. Hence the quotient map
\begin{equation}
\pi:E^{\times}_{\rm free}(d)\longrightarrow M^{\times}_{\rm free}
\end{equation}
is a connected regular covering with deck group $G_d$.
\end{prop}
\begin{proof}
Let $\widetilde G:=\Sp_2(\Z)\ltimes(d^{-1}\Z)^2$ act on $\hh\times\C$ by
$(\gamma,\bs{t}):(\tau,z)\mapsto\lr{\gamma.\tau,\,(z+t_1+t_2\tau)/(c\tau+d)}$, and let
$N:=\Gamma(d)\ltimes\Z^2\subset\widetilde G$, with
$E^{\times}(d)$ defined as in eq.~\eqref{eq:Ex}. %
The subgroup $N$ is normal in $\widetilde G$ as $\Gamma(d)$ is normal in $\Sp_2(\Z)$, and
conjugating $\gamma'\in\Gamma(d)$ by a torsion translation yields $\gamma'$ followed by a
translation by an element of $\Lambda_{\gamma'.\tau}$, since $\gamma'\equiv I \bmod d$.
Hence $\widetilde G/N\cong\Sp_2(\Z_d)\ltimes E[d]=G_d$ acts on $E^{\times}(d)$, which is
the action in the statement. For $d\geq3$ this action is effective: if $(g,\bs{v})$ acts
trivially, then $g$ acts trivially on $\Gamma(d)\backslash\hh$, so that $g=\pm I$ because a
generic $\tau$ has stabilizer $\{\pm I\}$ in $\Sp_2(\Z)$; on the fibers $(\pm I,\bs{v})$
acts as $z\mapsto\pm z+v$, which is the identity only for $g=I$ and $\bs{v}=0$.

For $d\geq3$, the space $E^{\times}(d)$ fibers over the connected surface $\Gamma(d)\backslash\hh$ with
connected fibers $E_\tau\setminus E_\tau[d]$ and is therefore connected. Since $G_d$ acts
holomorphically and effectively, the fixed-point set of each nonidentity element is a
proper complex-analytic subset, so that their union has real codimension at least two and
its complement $E^{\times}_{\rm free}(d)$ is path-connected. The free action of the finite
group $G_d$ on $E^{\times}_{\rm free}(d)$ is a covering-space action, and the quotient map
is a connected regular covering with deck group $G_d$~\cite[Prop.~1.40]{Hatcher}. %
For $d=2$, the element $-I\in\Gamma(2)$ acts trivially on $\hh$ and by $z\mapsto-z$ on the fibers, and it yields the only nontrivial isotropy of the $\Gamma(2)$-action~\cite{hain2014lectures}. Its fixed points are exactly $E_\tau[2]$,
which are removed in $E^{\times}(2)$, so that $E^{\times}(2)$ is a manifold with connected
coarse fibers $\lr{E_\tau\setminus E_\tau[2]}/\{\pm1\}$. The induced $G_2$-action is
effective, though no longer free on the translations, and the argument above applies
verbatim after restricting to $E^{\times}_{\rm free}(2)$.
\end{proof}
Concretely this statement says that, using unique path lifting, for a based loop
$\ell$ in $M^{\times}_{\rm free}$ and a chosen initial lift $\tilde x$, the lift of $\ell$
ends at $(g,\bs{v})\cdot\tilde x$ for a unique $(g,\bs{v})\in G_d$, and every element of
$G_d$ arises in this way. 
Relative to the level basis fixed by $\tilde x$, $g\in\Sp_2(\Z_d)$ records the symplectic Clifford action, while $\bs v\in E[d]$ records a logical Pauli displacement label (cf.\ Sec.~\ref{sec:Cliffords}).
For $d\geq 3$ and generic $\tau$, the nontrivial stabilizers come from $-I$ combined with a torsion translation. Their fixed points satisfy
\begin{equation}
2z\in d^{-1}\Lambda_\tau,
\qquad z\notin d^{-1}\Lambda_\tau.
\end{equation}
After quotienting by $E[d]$, their images are the three nonzero $2$-torsion classes of the syndrome torus. At these points the quotient has nontrivial isotropy, so the ordinary covering-space label is not available. Additional fixed points occur over the modular orbits of the elliptic curves with $\tau=i$ and $\tau=\rho$.
\begin{cor}[Positive-distance monodromy representatives]\label{cor:realization}
Let $d\geq 2$, fix $x_0\in M^{\times}_{\rm free}$, and choose a lift $\tilde x_0\in E^{\times}_{\rm free}(d)$. There exists $\varepsilon>0$ such that every $h\in G_d$ is the monodromy, relative to $\tilde x_0$, of a based loop at $x_0$. Each GKP code corresponding to points on the loop satisfies
\begin{equation}
\Delta_{\rm GKP}\geq\varepsilon.
\end{equation}
\end{cor}
\begin{proof}
By Proposition~\ref{prop:monodromy}, $E^{\times}_{\rm free}(d)$ is path-connected. For each $h\in G_d$, choose a path from $\tilde x_0$ to $h\cdot\tilde x_0$; its projection is a based loop with monodromy $h$. The union of these paths is compact because $G_d$ is finite. The $G_d$-invariant GKP distance is continuous and positive on this compact set, so it has a common positive lower bound $\varepsilon$, which also bounds the projected loops.
\end{proof}
Note that this corollary is purely a qualitative statement about the existence of a uniform distance lower bound.
A map $f:B\to M^{\times}_{\rm free}$ pulls this covering back to $B$. Globally, $M^{\times}$ remains an orbifold.
We summarize this property in fig.~\ref{fig:universal_family}. 

\subsection{Towards fiber bundle fault tolerance}\label{sec:fiber}

\begin{figure}
    \centering
    \includegraphics[scale=0.25]{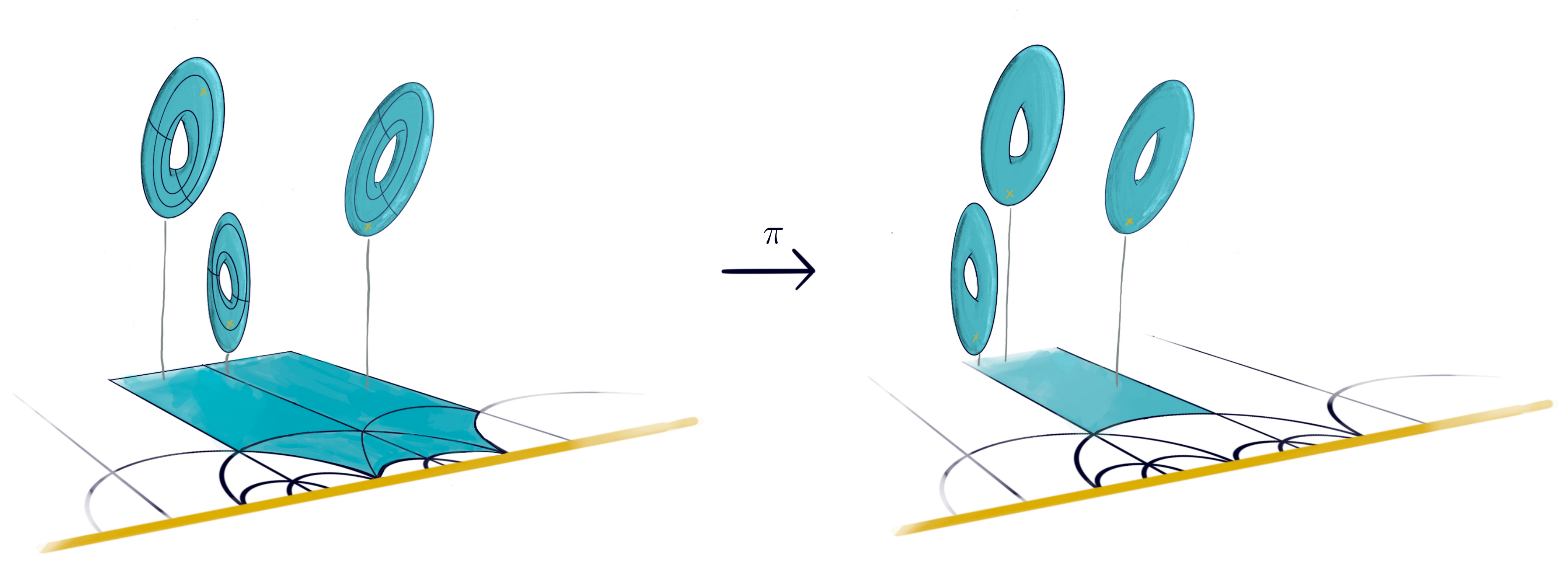}
    \caption{Schematic of the quotient in Eq.~\eqref{eq:EMspace}. For $d\geq3$, the left panel represents the level-marked family $E^\times(d)$ with fiber $E_\tau\setminus E_\tau[d]$. Quotienting by $E[d]$ forgets the torsion-displacement label and leaves a once-punctured syndrome torus. For $d=2$, both fiber descriptions additionally carry the quotient by $\{\pm1\}$. Quotienting subsequently by $\Sp_2(\Z_d)$ forgets the level basis. At points with trivial $G_d$-stabilizer, the combined map is a $G_d$-covering.}
    \label{fig:MX}
\end{figure}

Gottesman and Zhang~\cite{gottesman2017fibre} describe suitable families of quantum codes by vector bundles whose base points are code subspaces and whose fibers are the corresponding spaces of code states. Fault-tolerant unitary evolutions induce projective parallel transport, and logical gates are obtained as the resulting monodromies. We now explain the related covering-space structure furnished by the present construction.
We elaborate on this structure by taking successive quotients. For a fixed $\tau$, we first work on the elliptic curve before any residual automorphism quotient. The displacement coordinate in normalized units is taken modulo the stabilizer lattice, giving
\begin{equation}
E_\tau:=\C/\Lambda_\tau.
\end{equation}
With $y=\Im\tau$, the physical lattice $\CL_{\tau,d}=\sqrt{d/y}\Lambda_\tau$ has symplectic dual $\CL_{\tau,d}^{\perp}=d^{-1}\CL_{\tau,d}$. We use $z=\sqrt{y/d}\,w$ for a physical displacement $w$, so the stabilizer and dual lattices are represented in the $z$ coordinate by $\Lambda_\tau$ and $\Lambda_\tau^{\perp}:=d^{-1}\Lambda_\tau$, respectively.
A point $[z]_{\Lambda_\tau}\in E_\tau$ records the syndrome
$[z]_{\Lambda_\tau^\perp}$ together with a choice among its $d^2$
representatives in $E_\tau$. Any two such representatives differ by a
unique element of
\begin{equation}
E_\tau[d]
=
\Lambda_\tau^\perp/\Lambda_\tau
\cong(\Z/d\Z)^2,
\end{equation}
that is, by a unique logical Pauli displacement modulo phase.
Quotienting by translations in $E_\tau[d]$ therefore identifies all
representatives of the same syndrome and leaves the $z$ coordinate with
no logical-Pauli information.
The quotient chain
\begin{equation}
\C
\longrightarrow
E_\tau=\C/\Lambda_\tau
\longrightarrow
\Sigma_\tau:=E_\tau/E_\tau[d]
\cong
\C/\Lambda_\tau^\perp
\end{equation}
identifies all points that have the same syndrome but differ by a logical Pauli displacement. Consequently, the remaining coordinate $[z]_{\Lambda_\tau^\perp}\in\Sigma_\tau$ records the syndrome only and carries no logical-Pauli label. The $d^2$ points of $E_\tau[d]$ are the preimages of the single zero-syndrome point in $\Sigma_\tau$. Removing them therefore gives the regular covering
\begin{equation}
q_\tau:
E_\tau\setminus E_\tau[d]
\longrightarrow
\Sigma_\tau^\times
:=
\bigl(\C/\Lambda_\tau^\perp\bigr)\setminus\{0\},\label{eq:qtau}
\end{equation}
with deck group $E_\tau[d]$. 
The syndrome space $\Sigma_\tau^\times$ is a once-punctured torus with fundamental group $F_2$. Relative to the chosen level basis, a single winding around either fundamental torus cycle lifts to a path whose endpoints differ by one of the two logical Pauli generators. More generally, a loop with winding numbers $(m,n)$ has logical Pauli monodromy $X^m Z^n$ modulo phase, with the exponents reduced modulo $d$. A small loop around the removed zero represents their commutator up to inverse and conjugacy and has trivial monodromy in the abelian deck group $E_\tau[d]$, although it is itself noncontractible.

Since $E[d]$ is the normal translation subgroup of $G_d=\Sp_2(\Z_d)\ltimes E[d]$, the global quotient in Eq.~\eqref{eq:EMspace} can be performed in two stages,
\begin{equation}
E^\times(d)
\longrightarrow
E[d]\backslash E^\times(d)
\longrightarrow
\Sp_2(\Z_d)\backslash
\bigl(E[d]\backslash E^\times(d)\bigr)
\cong M^\times.
\end{equation}
The first arrow forgets the torsion-displacement label (logical Pauli label) by quotienting by the dual lattice in normalized coordinates, and the second forgets the symplectic level basis (logical Clifford label). For $d\geq3$, the translation action is free, while the remaining linear action can have fixed points. For $d=2$, the first quotient has fibers $\Sigma_\tau^\times/\{\pm1\}$, and translations can already have fixed points on $E^\times(2)$: for example, translation by $1/2$ fixes the class of $1/4$ modulo $\{\pm1\}$. In either case, the combined map is an ordinary covering after restriction to the subset with trivial $G_d$-stabilizer.

A based loop in $M^\times_{\rm free}$ lifts uniquely once its initial lift has been chosen. The deck transformation relating its endpoints is an element of $G_d$: its linear component records a symplectic Clifford action, and its translation component records a logical Pauli displacement label. The quotient therefore forgets these labels on the base while recovering their transformation through monodromy. This provides a covering-space model related to the Gottesman--Zhang picture, discussed further in Ref.~\cite{Burchards_GeometricGKP}.

In total, we obtain a complex orbifold of single-mode code shapes with nonzero syndrome, whose subset with trivial stabilizers carries a finite cover whose monodromy records symplectic Clifford actions and Pauli displacement labels. The zero-syndrome and finite-symmetry exclusions are distinct from the zero-distance degeneration; the positive-distance representatives of Corollary~\ref{cor:realization} refer to this restricted geometric construction.

\section{Discussion \& Outlook}
In this manuscript we have studied the structure and geometric model of Clifford-gate implementations for GKP codes. 

Motivated by the results of this study, we established a connection between GKP codes, the moduli space of symplectic lattices and algebraic curves. We pointed out how the configuration space of GKP codes, up to displacements, is given by the complement of a trefoil knot in the three sphere, $S^3-K$, and have shown that the knot defect corresponds to the limit of zero distance $\Delta_{\rm GKP}=0$ GKP codes.  For a primitive hyperbolic integral conjugacy class, the Rademacher symbol gives the linking number of the associated closed modular geodesic and, finally, our single-mode quotient provides a finite covering whose monodromy records symplectic Clifford actions and Pauli displacement labels on the subset without nontrivial symmetries, relating its geometry to a covering-space formulation of the proposal in Ref.~\cite{gottesman2017fibre}.
While we are only considering the simplest case of a single-mode $n=1$, the tools we develop hint at a general connection between the study of fiber bundle fault tolerance and the moduli spaces of (complex) algebraic varieties with additional structure. 
It will be the topic of future work to explore this connection in greater generality. 

Possible lines of future work, that we only developed vague hints for, involve the understanding of properties for GKP codes from Jacobians and how their complex structure and automorphisms manifest in a coding theoretic manner. Furthermore, as quantum states over Jacobians appear in certain conformal field theories~\cite{WittenAxelrod91} and Theta functions over symplectic lattices arise within the conformal bootstrap programme~\cite{Dymarsky_2021}, it is tempting to speculate that GKP codes allow us to construct further relationships between these avenues of research and quantum coding theory.

We have explained in depth how the limit of zero-distance GKP codes expresses itself as a 1-dimensional knot defect in the single-mode lattice-realization space. 
For multimode GKP codes, whose moduli spaces are higher-dimensional, we expect the structure of the relevant defect to be more complex and we expect their classification to yield further interesting implications for the geometric theory of fault tolerance.

Finally, rather than traversing the space of GKP codes by continuously parametrized Gaussian unitary operations, one may be able to take infinitesimal steps in this space by slightly varying the stabilizers measured. 
We anticipate that the understanding of the moduli space of GKP codes developed here also suggests a version of Floquet-based quantum error correction for GKP codes, where periodic measurement-cycles of varying generators can be used to implement logical gates of the GKP code, similar to the constructions presented in refs.~\cite{Hastings_2021, Vuillot_2019_codedeformation, davydova2023quantum, Townsend_Teague_2023, Kesselring_2024} for qubit-based codes.

We have demonstrated that tools from the theory of algebraic curves and modular forms play a significant role in the study of GKP codes and their fault tolerance properties. 
Since GKP codes also present an embedding of any qubit- or qudit-based stabilizer code into lattices, we hope that the present manuscript motivates further study of the general link between algebraic geometric formulations of quantum error correction and the theory of fault tolerance.
\medskip

Since the initial appearance of this manuscript, other expositions of the relationship between complex abelian varieties and GKP codes have appeared~\cite{Burchards_GeometricGKP, joseph2025generalizedgottesmankitaevpreskillstatesquantum, Conrad2024Fabulous, mayrand2026complexabelianvarietiesquantum}. These works highlight diverse aspects of the mathematical foundations of GKP codes and point towards a rich intersection of mathematical physics and quantum coding theory to be further explored.

\section*{Data availability}
The authors declare that the data supporting the findings of this study are available within the paper.

\acknowledgments
We acknowledge the use of AI (ChatGPT and Claude) for proofreading, suggesting changes (proposing proposition statements based on the previous prose style, flagging ambiguous wording and locating fixes in response to reviewer comments), and editing figures in preparation of the revised version of this manuscript.
We are grateful to %
V.\ Albert, 
J.\ Eisert, 
J.\ Franke,
J.\ Haferkamp, 
J.\ Ueki,
N.\ Walk and
J.\ Magdalena de la Fuente for many inspiring and helpful discussions and we are grateful to V.\ Albert for pointing out the fiber bundle framework for fault tolerance during discussions at the QIP conference in early 2022 as well as sharing thoughts on the topology of GKP codes. We also thank U. Chabaud for feedback on an early version of the manuscript.
JC thanks A. Grimsmo for introducing him to STF and for inviting him to a research visit at the AWS CQC.

JC and AB gratefully acknowledge support from the BMBF (RealistiQ, MUNIQC-Atoms, PhoQuant, QPIC-1, and QSolid), the DFG (CRC 183, project B04, on entangled states of matter), the Munich Quantum Valley (K-8), the ERC (DebuQC), Quantum Berlin as well as the Einstein Research Unit on quantum devices.

\bibliographystyle{unsrt}
\bibliography{Gardenbib}

@article{Dymarsky_2021,
   title={Quantum stabilizer codes, lattices, and {CFTs}},
   volume={2021},
   ISSN={1029-8479},
   url={http://dx.doi.org/10.1007/JHEP03(2021)160},
   DOI={10.1007/jhep03(2021)160},
   number={3},
   journal={Journal of High Energy Physics},
   publisher={Springer Science and Business Media LLC},
   author={Dymarsky, Anatoly and Shapere, Alfred},
   year={2021},
   month=Mar }

@article{WittenAxelrod91,
author = {Scott Axelrod and Steve Della Pietra and Edward Witten},
title = {{Geometric quantization of Chern-Simons gauge theory}},
volume = {33},
journal = {Journal of Differential Geometry},
number = {3},
publisher = {Lehigh University},
pages = {787 -- 902},
year = {1991},
doi = {10.4310/jdg/1214446565},
URL = {https://doi.org/10.4310/jdg/1214446565}
}

@misc{joseph2025generalizedgottesmankitaevpreskillstatesquantum,
      title={Generalized {Gottesman}-{Kitaev}-{Preskill} States on a Quantum Torus}, 
      author={Sijo K. Joseph and Sudhir Singh},
      year={2025},
      eprint={2509.18204},
      archivePrefix={arXiv},
      primaryClass={quant-ph},
      url={https://arxiv.org/abs/2509.18204}, 
}

@phdthesis{Conrad2024Fabulous,
  author  = {Conrad, Jonathan},
  title   = {The fabulous world of {GKP} codes},
  school  = {Freie Universit{\"a}t Berlin},
  address = {Berlin},
  year    = {2024},
  type    = {Doctoral dissertation},
  doi     = {10.17169/refubium-45505},
  url     = {https://doi.org/10.17169/refubium-45505}
}

@misc{mayrand2026complexabelianvarietiesquantum,
      title={Complex abelian varieties and quantum error correction: a mathematical framework for {GKP} codes}, 
      author={Maxence Mayrand and Baptiste Royer},
      year={2026},
      eprint={2605.28784},
      archivePrefix={arXiv},
      primaryClass={math.AG},
      url={https://arxiv.org/abs/2605.28784}, 
}

@article{Vuillot_2019_codedeformation,
   title={Code deformation and lattice surgery are gauge fixing},
   volume={21},
   ISSN={1367-2630},
   url={http://dx.doi.org/10.1088/1367-2630/ab0199},
   DOI={10.1088/1367-2630/ab0199},
   number={3},
   journal={New Journal of Physics},
   publisher={IOP Publishing},
   author={Vuillot, Christophe and Lao, Lingling and Criger, Ben and Garc{\'i}a Almud{\'e}ver, Carmen and Bertels, Koen and Terhal, Barbara M.},
   year={2019},
   month=mar, pages={033028} }

@misc{davydova2023quantum,
      title={Quantum computation from dynamic automorphism codes}, 
      author={Margarita Davydova and Nathanan Tantivasadakarn and Shankar Balasubramanian and David Aasen},
      year={2023},
      eprint={2307.10353},
      archivePrefix={arXiv},
      primaryClass={quant-ph}
}

@article{Hastings_2021,
   title={Dynamically Generated Logical Qubits},
   volume={5},
   ISSN={2521-327X},
   url={http://dx.doi.org/10.22331/q-2021-10-19-564},
   DOI={10.22331/q-2021-10-19-564},
   journal={Quantum},
   publisher={Verein zur Forderung des Open Access Publizierens in den Quantenwissenschaften},
   author={Hastings, Matthew B. and Haah, Jeongwan},
   year={2021},
   month=oct, pages={564} }

@article{Townsend_Teague_2023,
   title={Floquetifying the Colour Code},
   volume={384},
   ISSN={2075-2180},
   url={http://dx.doi.org/10.4204/EPTCS.384.14},
   DOI={10.4204/eptcs.384.14},
   journal={Electronic Proceedings in Theoretical Computer Science},
   publisher={Open Publishing Association},
   author={Townsend-Teague, Alex and Magdalena de la Fuente, Julio and Kesselring, Markus},
   year={2023},
   month=aug, pages={265–303} }

@article{Kesselring_2024,
   title={Anyon Condensation and the Color Code},
   volume={5},
   ISSN={2691-3399},
   url={http://dx.doi.org/10.1103/PRXQuantum.5.010342},
   DOI={10.1103/prxquantum.5.010342},
   number={1},
   journal={PRX Quantum},
   publisher={American Physical Society (APS)},
   author={Kesselring, Markus S. and Magdalena de la Fuente, Julio C. and Thomsen, Felix and Eisert, Jens and Bartlett, Stephen D. and Brown, Benjamin J.},
   year={2024},
   month=mar }

@book{Gannon2023,
  title = {Moonshine beyond the {Monster}: The Bridge Connecting Algebra,  Modular Forms and Physics},
  ISBN = {9781009401555},
  url = {http://dx.doi.org/10.1017/9781009401548},
  DOI = {10.1017/9781009401548},
  publisher = {Cambridge University Press},
  author = {Gannon,  Terry},
  year = {2023},
  month = jul 
}

@Inbook{Bobenko2011,
author="Bobenko, Alexander I.",
title="Introduction to Compact {Riemann} Surfaces",
bookTitle="Computational Approach to {Riemann} Surfaces",
year="2011",
publisher="Springer Berlin Heidelberg",
address="Berlin, Heidelberg",
pages="3--64",
isbn="978-3-642-17413-1",
doi="10.1007/978-3-642-17413-1_1",
url="https://doi.org/10.1007/978-3-642-17413-1_1",
  crossref = "bobenko2011computational",
}

@book{bobenko2011computational,
  title={Computational Approach to {Riemann} Surfaces},
  author={Bobenko, A.I. and Klein, C.},
  number={Bd. 2013},
  isbn={9783642174124},
  lccn={2011920817},
  series={Computational Approach to Riemann Surfaces},
  url={https://books.google.de/books?id=53QpxVeFzcEC},
  year={2011},
  publisher={Springer}
}

@misc{gottesman2022opportunities,
      title={Opportunities and Challenges in Fault-Tolerant Quantum Computation}, 
      author={Daniel Gottesman},
      year={2022},
      eprint={2210.15844},
      archivePrefix={arXiv},
      primaryClass={quant-ph}
}

@book{Milnor+1972,
url = {https://doi.org/10.1515/9781400881796},
title = {Introduction to Algebraic K-Theory. (AM-72), Volume 72},
author = {John Milnor},
publisher = {Princeton University Press},
address = {Princeton},
doi = {doi:10.1515/9781400881796},
isbn = {9781400881796},
year = {1972},
lastchecked = {2024-05-06}
}

@article{Duke,
author = {W. Duke and {\"O}. Imamo{\=g}lu and {\'A}. T{\'o}th},
title = {{Modular cocycles and linking numbers}},
volume = {166},
journal = {Duke Mathematical Journal},
number = {6},
publisher = {Duke University Press},
pages = {1179 -- 1210},
year = {2017},
doi = {10.1215/00127094-3793032},
URL = {https://doi.org/10.1215/00127094-3793032}
}

@article{Ueki,
   title={Modular knots, automorphic forms, and the {Rademacher} symbols for triangle groups},
   volume={10},
   ISSN={2197-9847},
   url={http://dx.doi.org/10.1007/s40687-022-00366-8},
   DOI={10.1007/s40687-022-00366-8},
   number={1},
   journal={Research in the Mathematical Sciences},
   publisher={Springer Science and Business Media LLC},
   author={Matsusaka, Toshiki and Ueki, Jun},
   year={2022},
   month=dec }

@misc{AMSGhys,
title={Lorenz and Modular Flows: A Visual Introduction},
author={Ghys,  E.},
url={https://www.ams.org/publicoutreach/feature-column/fcarc-lorenz},
year={2006}
}

@Inbook{Zagier2008,
author="Zagier, Don",
title="Elliptic Modular Forms and Their Applications",
bookTitle="The 1-2-3 of Modular Forms: Lectures at a Summer School in Nordfjordeid, Norway",
year="2008",
publisher="Springer Berlin Heidelberg",
address="Berlin, Heidelberg",
pages="1--103",
isbn="978-3-540-74119-0",
doi="10.1007/978-3-540-74119-0_1",
url="https://doi.org/10.1007/978-3-540-74119-0_1",
crossret="ranestad20081"
}

@book{Silverman2009,
  title = {The Arithmetic of Elliptic Curves},
  ISBN = {9780387094946},
  ISSN = {2197-5612},
  url = {http://dx.doi.org/10.1007/978-0-387-09494-6},
  DOI = {10.1007/978-0-387-09494-6},
  journal = {Graduate Texts in Mathematics},
  publisher = {Springer New York},
  author = {Silverman,  Joseph H.},
  year = {2009}
}

@misc{hain2014lectures,
      title={Lectures on Moduli Spaces of Elliptic Curves}, 
      author={Richard Hain},
      year={2014},
      eprint={0812.1803},
      archivePrefix={arXiv},
      primaryClass={math.AG}
}

@misc{grushevsky2010schottky,
      title={The {Schottky} Problem}, 
      author={Samuel Grushevsky},
      year={2010},
      eprint={1009.0369},
      archivePrefix={arXiv},
      primaryClass={math.AG}
}

@book{Griffiths1989,
  title = {Introduction to Algebraic Curves},
  ISBN = {9781470444907},
  ISSN = {2472-5137},
  url = {http://dx.doi.org/10.1090/mmono/076},
  DOI = {10.1090/mmono/076},
  journal = {Translations of Mathematical
                        Monographs},
  publisher = {American Mathematical
                    Society},
  author = {Griffiths,  Phillip},
  year = {1989},
  month = dec 
}

@incollection {Berge,
    AUTHOR = {Berg{\'e}, Anne-Marie},
     TITLE = {Symplectic lattices},
 BOOKTITLE = {Quadratic Forms and Their Applications},
    SERIES = {Contemporary Mathematics},
    VOLUME = {272},
     PAGES = {9--22},
 PUBLISHER = {American Mathematical Society},
   ADDRESS = {Providence, RI},
      YEAR = {2000},
      ISBN = {978-0-8218-2779-6},
      CROSSREF = {BayerFluckigerLewisRanicki2000}
}

@proceedings{BayerFluckigerLewisRanicki2000,
  editor    = {Bayer-Fluckiger, Eva and Lewis, David and Ranicki, Andrew},
  title     = {Quadratic Forms and Their Applications},
  series    = {Contemporary Mathematics},
  volume    = {272},
  publisher = {American Mathematical Society},
  year      = {2000},
  isbn      = {9780821827796},
  doi       = {10.1090/conm/272},
  url       = {https://bookstore.ams.org/CONM/272}
}

@article{Burchards_GeometricGKP,
    title = {Fiber {B}undle {F}ault {T}olerance of {GKP} {C}odes},
    author = {Burchards, Ansgar G. and Flammia, Steven T. and Conrad, Jonathan},
    journal = {Quantum},
    volume = {9},
    pages = {1899},
    year = {2025},
    doi = {10.22331/q-2025-10-29-1899},
    url = {https://doi.org/10.22331/q-2025-10-29-1899}
}

@article{Albert_2018,
  title = {Performance and structure of single-mode bosonic codes},
  author = {Albert, Victor V. and Noh, Kyungjoo and Duivenvoorden, Kasper and Young, Dylan J. and Brierley, R. T. and Reinhold, Philip and Vuillot, Christophe and Li, Linshu and Shen, Chao and Girvin, S. M. and Terhal, Barbara M. and Jiang, Liang},
  journal = {Phys. Rev. A},
  volume = {97},
  issue = {3},
  pages = {032346},
  numpages = {30},
  year = {2018},
  month = {Mar},
  publisher = {American Physical Society},
  doi = {10.1103/PhysRevA.97.032346},
  url = {https://link.aps.org/doi/10.1103/PhysRevA.97.032346}
}

@article{Sivak_2023,
	doi = {10.1038/s41586-023-05782-6},
  
	url = {https://doi.org/10.1038\%2Fs41586-023-05782-6},
  
	year = 2023,
	month = {mar},
  
	publisher = {Springer Science and Business Media {LLC}
},
  
	volume = {616},
  
	number = {7955},
  
	pages = {50--55},
  
	author = {V. V. Sivak and A. Eickbusch and B. Royer and S. Singh and I. Tsioutsios and S. Ganjam and A. Miano and B. L. Brock and A. Z. Ding and L. Frunzio and S. M. Girvin and R. J. Schoelkopf and M. H. Devoret},
  
	title = {Real-time quantum error correction beyond break-even},
  
	journal = {Nature}
}

@article{Seifert,
author = {H. Seifert},
title = {{Topologie Dreidimensionaler Gefaserter Räume}},
volume = {60},
journal = {Acta Mathematica},
publisher = {Institut Mittag-Leffler},
pages = {147 -- 238},
year = {1933},
doi = {10.1007/BF02398271},
URL = {https://doi.org/10.1007/BF02398271}
}

@article{Ghys,
author = {Ghys, {\'E}tienne},
year = {2006},
month = {01},
pages = {},
title = {Knots and dynamics},
volume = {1},
isbn = {978-3-03719-022-7},
journal = {Proceedings of the International Congress of Mathematicians, Vol. 1, 2006-01-01, ISBN 978-3-03719-022-7, pags. 247-277},
doi = {10.4171/022-1/11}
}

@misc{conrad2023good,
      title={Good {Gottesman}-{Kitaev}-{Preskill} codes from the {NTRU} cryptosystem}, 
      author={Jonathan Conrad and Jens Eisert and Jean-Pierre Seifert},
      year={2023},
      eprint={2303.02432},
      archivePrefix={arXiv},
      primaryClass={quant-ph}
}

@book{Bourbaki9,
title = {Alg{\'{e}}bre},
author={Bourbaki},
ISBN={9783540353386}, 
url={http://dx.doi.org/10.1007/978-3-540-35339-3}, 
DOI={10.1007/978-3-540-35339-3}, 
publisher={Springer Berlin Heidelberg}, 
year={2007} }

@book{Birkenhake_2004,
	doi = {10.1007/978-3-662-06307-1},
	url = {https://doi.org/10.1007%2F978-3-662-06307-1},
	year = {2004},
	publisher = {Springer Berlin Heidelberg},
	author = {Christina Birkenhake and Herbert Lange},
	title = {Complex {Abelian} Varieties}
}

@misc{gottesman2017fibre,
      title={Fibre bundle framework for unitary quantum fault tolerance}, 
      author={Daniel Gottesman and Lucy Liuxuan Zhang},
      year={2017},
      eprint={1309.7062},
      archivePrefix={arXiv},
      primaryClass={quant-ph},
      url={https://arxiv.org/abs/1309.7062}, 
}

@book{FarbMargalit+2012,
url = {https://doi.org/10.1515/9781400839049},
title = {A Primer on Mapping Class Groups (PMS-49)},
author = {Benson Farb and Dan Margalit},
publisher = {Princeton University Press},
address = {Princeton},
doi = {doi:10.1515/9781400839049},
isbn = {9781400839049},
year = {2012},
lastchecked = {2023-08-03}
}

@book{omeara_symplectic_1978,
	series = {Mathematical {Surveys} and {Monographs}},
	title = {Symplectic {Groups}},
	isbn = {978-0-8218-1516-8},
	url = {https://books.google.de/books?id=BWHyBwAAQBAJ},
	publisher = {American Mathematical Society},
	author = {O'Meara, O.T.},
	year = {1978},
	lccn = {78019101},
}

@article{AllGaussian,
  title = {All-{Gaussian} Universality and Fault Tolerance with the {Gottesman}-{Kitaev}-{Preskill} Code},
  author = {Baragiola, Ben Q. and Pantaleoni, Giacomo and Alexander, Rafael N. and Karanjai, Angela and Menicucci, Nicolas C.},
  journal = {Phys. Rev. Lett.},
  volume = {123},
  issue = {20},
  pages = {200502},
  numpages = {6},
  year = {2019},
  month = {Nov},
  publisher = {American Physical Society},
  doi = {10.1103/PhysRevLett.123.200502},
  url = {https://link.aps.org/doi/10.1103/PhysRevLett.123.200502}
}

@article{GCB,
  title = {Quantum Computing with Rotation-Symmetric Bosonic Codes},
  author = {Grimsmo, Arne L. and Combes, Joshua and Baragiola, Ben Q.},
  journal = {Phys. Rev. X},
  volume = {10},
  issue = {1},
  pages = {011058},
  numpages = {32},
  year = {2020},
  month = {Mar},
  publisher = {American Physical Society},
  doi = {10.1103/PhysRevX.10.011058},
  url = {https://link.aps.org/doi/10.1103/PhysRevX.10.011058}
}

@misc{Conrad_lectures,
  author = {Conrad, Keith},
  title = {Lecture notes},
  howpublished = "\url{https://kconrad.math.uconn.edu/blurbs/}",
  year = {2024}, 
  note = "[Online; accessed 06-May-2024]"
}

@article{Royer_2022,
  title = {Encoding Qubits in Multimode Grid States},
  author = {Royer, B. and Singh, S. and Girvin, S. M.},
  journal = {PRX Quantum},
  volume = {3},
  issue = {1},
  pages = {010335},
  numpages = {40},
  year = {2022},
  month = {Mar},
  publisher = {American Physical Society},
  doi = {10.1103/PRXQuantum.3.010335},
  url = {https://link.aps.org/doi/10.1103/PRXQuantum.3.010335}
}

@article{Schmidt_2022,
  title = {{Quantum error correction with higher {Gottesman}-{Kitaev}-{Preskill} codes: Minimal measurements and linear optics}},
  author = {Schmidt, F. and van Loock, P.},
  journal = {Phys. Rev. A},
  volume = {105},
  issue = {4},
  pages = {042427},
  numpages = {20},
  year = {2022},
  month = {Apr},
  publisher = {American Physical Society},
  doi = {10.1103/PhysRevA.105.042427},
  url = {https://link.aps.org/doi/10.1103/PhysRevA.105.042427}
}

@article{Conrad_2022,
  doi = {10.22331/q-2022-02-10-648},
  url = {https://doi.org/10.22331/q-2022-02-10-648},
  title = {Gottesman-{K}itaev-{P}reskill codes: {A} lattice perspective},
  author = {Conrad, J. and Eisert, J. and Arzani, F.},
  journal = {{Quantum}},
  issn = {2521-327X},
  publisher = {{Verein zur F{\"{o}}rderung des Open Access Publizierens in den Quantenwissenschaften}},
  volume = {6},
  pages = {648},
  month = feb,
  year = {2022}
}

@article{Bravyi_2005,
	doi = {10.1103/physreva.71.022316},
  
	url = {https://doi.org/10.1103%2Fphysreva.71.022316},
  
	year = 2005,
	month = {feb},
  
	publisher = {American Physical Society ({APS})},
  
	volume = {71},
  
	number = {2},
  
	author = {Sergey Bravyi and Alexei Kitaev},
  
	title = {Universal quantum computation with ideal {Clifford} gates and noisy ancillas},
  
	journal = {Physical Review A}
}

@article{Duivenvoorden_Sensor,
  title = {Single-mode displacement sensor},
  author = {Duivenvoorden, K. and Terhal, B. M. and Weigand, D.},
  journal = {Phys. Rev. A},
  volume = {95},
  pages = {012305},
  numpages = {15},
  year = {2017},
  publisher = {American Physical Society},
  doi = {10.1103/PhysRevA.95.012305},
  url = {https://link.aps.org/doi/10.1103/PhysRevA.95.012305}
}

@article{Weedbrook_2012,
   title={Gaussian quantum information},
   volume={84},
   ISSN={1539-0756},
   url={http://dx.doi.org/10.1103/RevModPhys.84.621},
   DOI={10.1103/revmodphys.84.621},
   number={2},
   journal={Rev. Mod. Phys.},
   publisher={American Physical Society (APS)},
   author={C.  Weedbrook and S.  Pirandola and R.  Garc{\'i}a-Patr{\'o}n  and N. J.  Cerf  and T.C.  Ralph  and  J.H.  Shapiro  and S.  Lloyd},
   year={2012},
   month={May},
   pages={621–669}
}

@ARTICLE{Noh_Capacity,
  author={K.  Noh and V. V.  Albert and L.  Jiang},
  journal={IEEE Trans. Inf. Th.}, 
  title={{Quantum capacity bounds of {Gaussian} thermal loss channels and achievable rates with {Gottesman}-{Kitaev}-{Preskill} codes}}, 
  year={2019},
  volume={65},
  pages={2563-2582},
  doi={10.1109/TIT.2018.2873764}}

@article{Fluehmann_2019,
	Author = {Fl{\"u}hmann, C. and Nguyen, T. L. and Marinelli, M. and Negnevitsky, V. and Mehta, K. and Home, J. P.},
	Doi = {10.1038/s41586-019-0960-6},
	Issn = {1476-4687},
	Journal = {Nature},
	Pages = {513--517},
	Publisher = {Springer Science and Business Media LLC},
	Title = {Encoding a qubit in a trapped-ion mechanical oscillator},
	Url = {http://dx.doi.org/10.1038/s41586-019-0960-6},
	Volume = {566},
	Year = {2019}}

@article{Campagne_Ibarcq_2020,
	Author = {Campagne-Ibarcq, P. and Eickbusch, A. and Touzard, S. and Zalys-Geller, E. and Frattini, N. E. and Sivak, V. V. and Reinhold, P. and Puri, S. and Shankar, S. and Schoelkopf, R. J. and others},
	Doi = {10.1038/s41586-020-2603-3},
	Issn = {1476-4687},
	Journal = {Nature},
	Pages = {368--372},
	Publisher = {Springer Science and Business Media LLC},
	Title = {Quantum error correction of a qubit encoded in grid states of an oscillator},
	Url = {http://dx.doi.org/10.1038/s41586-020-2603-3},
	Volume = {584},
	Year = {2020}}

@article{GKP,
  title = {Encoding a qubit in an oscillator},
  author = {Gottesman, D.  and Kitaev, A.  and Preskill, J.},
  journal = {Phys. Rev. A},
  volume = {64},
  pages = {012310},
  numpages = {21},
  year = {2001},
  publisher = {American Physical Society},
  doi = {10.1103/PhysRevA.64.012310},
  url = {https://link.aps.org/doi/10.1103/PhysRevA.64.012310}
}

@phdthesis{Harrington_Thesis,
  author       = {Harrington, J. W.}, 
  title        = {Analysis of quantum error-correcting codes: Symplectic lattice codes and toric codes},
  school       = {California Institute of Technology},
  year         = {2004},
 doi =	{10.7907/AHMQ-EG82},
 url={https://resolver.caltech.edu/CaltechETD:etd-05122004-113132}
}

@article{Kitaev_2003,
   title={Fault-tolerant quantum computation by anyons},
   volume={303},
   url={http://dx.doi.org/10.1016/S0003-4916(02)00018-0},
   DOI={10.1016/s0003-4916(02)00018-0},
   journal={Ann. Phys.},
   publisher={Elsevier BV},
   author={Kitaev, A. Y.},
   year={2003},
   pages={2–30}
}

@article{Terhal_2020,
	doi = {10.1088/2058-9565/ab98a5},
	url = {https://doi.org/10.1088/2058-9565/ab98a5},
	year = 2020,
	publisher = {{IOP} Publishing},
	volume = {5},
	pages = {043001},
	author = {B. M. Terhal and J. Conrad and C. Vuillot},
	title = {Towards scalable bosonic quantum error correction},
	journal = {Quantum Science and Technology}
}

@article{Sarnak1994,
author = {Sarnak, P. and Buser, P.},
journal = {Inventiones mathematicae},
pages = {27-56},
title = {{On the period matrix of a {Riemann} surface of large genus (with an Appendix by J. H. {Conway} and N. J. A. {Sloane})}},
url = {http://eudml.org/doc/144207},
volume = {117},
year = {1994},
}

@article{HarringtonPreskill,
  title = {{Achievable rates for the {Gaussian} quantum channel}},
  author = {Harrington, J. and Preskill, J.},
  journal = {Phys. Rev. A},
  volume = {64},
  issue = {6},
  pages = {062301},
  numpages = {10},
  year = {2001},
  publisher = {American Physical Society},
  doi = {10.1103/PhysRevA.64.062301},
  url = {https://link.aps.org/doi/10.1103/PhysRevA.64.062301}
}

@misc{GottesmanQECtalk,
    author = {Daniel Gottesman},
    title = {{The Definition(s) of Fault Tolerance}},
    note = {Invited talk given at 4th International Conference on Quantum Error Correction; available at \url{https://youtu.be/FMXFNCIaF3k}},
    year = {2017},
}

@book{Hatcher,
    author = {Allen Hatcher},
    title = {Algebraic Topology},
    publisher = {Cambridge University Press},
    address = {Cambridge},
    year = {2002},
}

\appendix
\section{More on quantum harmonic oscillators}\label{app:QHO}
Bosonic quantum error correction studies the robust embedding of discrete quantum information into a system of multiple quantum harmonic oscillators (QHO), each of which  can be described by an infinite-dimensional Hilbert space $\CH=\mathrm{span}\lrc{\ket{n}}_{n=0}^{\infty}$ where $\ket{n}$ denote the Fock states whose labels correspond to the eigenvalues of the number operator $\hat{n}=\hat{a}^{\dagger}\hat{a}$ and $\hat{a}=(\hat{q}+i\hat{p})/\sqrt{2}$ denotes the annihilation operator. $\hat{q}$ and $\hat{p}$ are the position and momentum operators whose improper eigenstates yield a basis for the underlying Hilbert space. The associated phase space inherits a non-trivial geometry from the canonical commutation relations $\lrq{\hat{q}, \hat{p}}=i$ (we will set $\hbar=1$ throughout) and is most naturally studied in the Heisenberg frame, i.e. in terms of the transformation behaviour of operators on this space.  On a system of $n$ QHOs -- which we will refer to as having $n$ \textit{modes} --  we define a generalized quadrature operator  $\bs{\hat{x}}=\lr{\hat{q}_1,\ldots,\hat{q}_n,\hat{p}_1,\ldots,\hat{p}_n}^T$ so that $\lrq{\hat{x}_i, \hat{x}_j}=iJ_{ij}$ where 
\begin{equation}
J=\begin{pmatrix}
0 & I_n \\ -I_n & 0
\end{pmatrix}
\end{equation} is the anti-symmetric symplectic form and $I_n$ denotes the $n \times n$ identity matrix.

Analogous to the Pauli-operators for qubit-systems, the Heisenberg-Weyl operators for this infinite dimensional Hilbert space are given by displacement operators
\begin{align}
D\lr{\bs{\xi}} \coloneqq \exp\left\{-i \sqrt{2\pi} \bs{\xi}^T J \bs{\hat{x}}\right\},
\end{align}
which form a basis for operators and  satisfy the orthogonality relation $\Tr\lrq{D^{\dagger}\lr{\bs{\xi}}D\lr{\bs{\eta}}}=\delta^{(2n)}\lr{\bs{\xi}-\bs{\eta}}$, therefore states can be represented by their Wigner function
\begin{equation}
W_{\rho}\lr{\bs{x}}
\coloneqq \int_{\mathbb{R}^{2n}} d\bs{\eta}\, e^{-i2\pi \bs{x}^T J \bs{\eta}} \Tr\lrq{D\lr{\bs{\eta}} \rho}.
\end{equation}

Displacement operators represent the unitary time evolution induced by Hamiltonians linear in the quadrature operators that implement the transformation 
\begin{equation}
D\lr{\bs{\xi}}^{\dagger} \bs{\hat{x}}D\lr{\bs{\xi}}=\bs{\hat{x}} + \sqrt{2\pi}\bs{\xi}
\end{equation}
 and commute and multiply as
\begin{align}
D\lr{\bs{\xi}}D\lr{\bs{\eta}}&=e^{-i\pi \bs{\xi}^TJ\bs{\eta}}D\lr{\bs{\xi}+\bs{\eta}}, \\
&=e^{-i2\pi \bs{\xi}^TJ\bs{\eta}}D\lr{\bs{\eta}}D\lr{\bs{\xi}}.\label{eq:displacement_commutator}
\end{align}
It is these properties that make them a natural set to choose stabilizer groups from.

Unitary evolution via Hamiltonians strictly quadratic in the quadrature operators implements symplectic transformations
\begin{align}
U_S\coloneqq e^{-\frac{i}{2}\bs{\hat{x}}^T C \bs{\hat{x}}},\; C=C^T, \\
U_S^{\dagger}\bs{\hat{x}}U_S=S\bs{\hat{x}},\; S=e^{JC}, \label{eq:sympH}
\end{align}
where $S\in \Sp_{2n}\lr{\R}=\lrc{S\in \R^{2n\times 2n} : S^TJS=J}$ is a symplectic matrix which follows from unitarity of $U_S$ and we have 
\begin{equation}
U_{S}D\lr{\bs{\xi}}U_{S}^{\dagger}=D\lr{S\bs{\xi}},
\end{equation}
and hence
\begin{equation}
W_{U_S\rho U_S^{\dagger}}\lr{\bs{x}}=W_{\rho}\lr{S^{-1}\bs{x}}.
\end{equation}

An important symplectic transformation is the \textit{symplectic transvection}
\begin{equation}
t_{\bs{\alpha}}\coloneqq I+\bs{\alpha}\bs{\alpha}^TJ,
\end{equation}
generated by the matrix $C=J^T\bs{\alpha}\bs{\alpha}^TJ$ so that the corresponding Hamiltonian takes the form $H=\frac{1}{2}(\bs{\alpha}^TJ\bs{\hat{x}})^2$. 
Symplectic transvections close under conjugation 
\begin{equation}
t_{\bs{\beta}}t_{\bs{\alpha}}t_{\bs{\beta}}^{-1}=t_{t_{\bs{\beta}}\bs{\alpha}}
\end{equation}
and are the homological actions of Dehn twists. These transvections generate the symplectic image of the mapping class group (not the full mapping class group), whose Torelli subgroup acts trivially on first homology, see refs.~\cite{omeara_symplectic_1978} and~\cite[Prop.~6.3, Thm.~6.4 and Sec.~6.5]{FarbMargalit+2012}.

Displacement operators also admit the following complex parametrization
\begin{align}
D_c\lr{\bs{\gamma}}\coloneqq\exp\lrc{\sqrt{\pi}\lr{\bs{\gamma}^T\bs{\hat{a}^{\dagger}}-\bs{\gamma}^{* T}\bs{\hat{a}}}} =D\lr{\bs{\xi}},
\end{align}
where the equivalent real parameter is $\bs{\xi}=\Re\lr{\bs{\gamma}}\oplus \Im\lr{\bs{\gamma}}\in \R^{2n}$ and $\bs{\hat{a}}=(\hat{a}_1,\hdots, \hat{a}_n), $ $\bs{\hat{a}^{\dagger}}=(\hat{a}^{\dagger}_1,\hdots, \hat{a}^{\dagger}_n)$ define the generalized annihilation and creation operators. 

In this parametrization the displacement operator acts as
\begin{equation}
D_c\lr{\bs{\gamma}}^{\dagger}\bs{\hat{a}}D_c\lr{\bs{\gamma}}=\bs{\hat{a}}+\sqrt{\pi}\bs{\gamma}.
\end{equation}
and the commutation relation is given by
\begin{equation}
D_c\lr{\bs{\gamma}}D_c\lr{\bs{\delta}}= e^{-i2\pi\Im\lr{ \bs{\gamma}^{\dagger}\bs{\delta}}} D_c\lr{\bs{\delta}}D_c\lr{\bs{\gamma}}
\end{equation}
where the symplectic form $\omega\lr{\bs{\gamma},  \bs{\delta}}=\Im\lr{ \bs{\gamma}^{\dagger}\bs{\delta}}$ is a skew-symmetric function inherited from the hermitian form $H\lr{\bs{\gamma}, \bs{\delta}}= \bs{\gamma}^{\dagger}\bs{\delta}$ as its imaginary part.

\section{GKP Clifford gates}\label{app:Cliff}

\begin{lem}\label{lem:integral_rep}
    Given a symplectically integral lattice $\CL \subseteq\CL^{\perp}$ with symplectic Gram matrix (symplectic form) $A=J_2\otimes D$, $D=\mathrm{diag}\lr{d_1,\hdots, d_n},$ we have that
    $\Aut^S\lr{\mathcal{L}}$ is equivalently specified by the integral representation
    \begin{equation}
        \Sp_{2n}^D\lr{\Z}=\{U\in \GL_{2n}\lr{\Z}:\; UAU^T=A\}.
    \end{equation}
\end{lem}

\proof
Let $M_D=D\oplus I_n$, so $A=M_DJM_D^T$. For $U\in\Sp_{2n}^D(\Z)$, define $S_U^T=M_D^{-1}UM_D$. Then
\begin{equation}
S_U^TJS_U=M_D^{-1}UAU^TM_D^{-T}=M_D^{-1}AM_D^{-T}=J,
\end{equation}
so $S_U\in\Sp_{2n}(\R)$ and $UM_D=M_DS_U^T$. For a general basis $M=M_DS_0^T$, the corresponding real automorphism is $S_0S_US_0^{-1}$. Conversely, if $UM=MS^T$ with $S$ symplectic, then $UAU^T=MS^TJSM^T=A$.
\endproof

\begin{cor}[Dual-lattice invariance]\label{cor:dual-aut}
For every full-rank lattice $\CL\subset\R^{2n}$,
\begin{equation}
\Aut^S\lr{\CL}=\Aut^S\lr{\CL^{\perp}}.
\end{equation}
\end{cor}

\proof
Let $S\in\Aut^S(\CL)$, $\eta\in\CL^{\perp}$, and $\xi\in\CL$. Since $S^TJ=JS^{-1}$ and $S^{-1}\xi\in\CL$,
\begin{equation}
(S\eta)^TJ\xi=\eta^TS^TJ\xi=\eta^TJS^{-1}\xi\in\Z.
\end{equation}
Thus $S\CL^{\perp}\subseteq\CL^{\perp}$, and applying the same argument to $S^{-1}$ gives equality. Hence $\Aut^S(\CL)\subseteq\Aut^S(\CL^{\perp})$. Applying this inclusion to $\CL^{\perp}$ and using $(\CL^{\perp})^{\perp}=\CL$ gives the reverse inclusion.
\endproof

Similarly, for the symplectic orthogonal automorphism group $\Aut^{SO}(\CL)=\Aut^{S}(\CL)\cap O_{2n}(\R)$ one has $\Aut^{SO}(\CL)=\Aut^{SO}(\CL^{\perp})$; its integral representation is the subgroup of $\Sp^D_{2n}(\Z)$ preserving in addition the Euclidean Gram matrix $G=MM^T$. 
Since $\Aut^S\lr{\CL^{\perp}}=\Aut^S\lr{\CL}$ by the corollary, every symplectic automorphism preserves both lattices and hence descends to an automorphism of $\CL^{\perp}/\CL$. For \textit{scaled GKP codes} (see ref.~\cite{Conrad_2022}), $D=dI_n$, the integral representation is all of $\Sp_{2n}(\Z)$, and reduction modulo $d$ maps it onto $\Sp_{2n}(\Z_d)$.

\section{Magic states}\label{app:Magic}
The elements of the symplectic orthogonal automorphism group $\Aut^{SO}\lr{\CL^{\perp}}$ of GKP codes have a special application in that they can give rise to magic states, as illustrated below.
Let $\ket{0}^{\otimes n}$ be the $n$-mode vacuum state. 
The vacuum state, with $\hat{U}_S\ket{\rm vac}^{\otimes n}=\ket{\rm vac}^{\otimes n}\, \forall S\in \Sp_{2n}(\R)\cap \SO_{2n}(\R)$ is rotation symmetric and arguably the simplest state to prepare. 
Further let 
\begin{equation}
\Pi_{M}\coloneqq\sum_{\bs{\xi}\in \CL\lr{M}} e^{i\phi_M\lr{\bs{\xi}}} D\lr{\bs{\xi}} 
\end{equation}
be the formal code-space projector of a GKP code with generator $M$, where the phases $\phi_M(\bs{\xi})$ are the appropriate Heisenberg cocycles, cf.\ ref.~\cite{Conrad_2022}. In the case of a scaled GKP code where the symplectic Gram matrix $A$ has only even entries we may choose trivial stabilizer phases $\phi_M\lr{\bs{\xi}} =0 \mod 2\pi$, so that we simply write $\Pi_{\CL}$ and for $\hat{U}_S$ the Gaussian unitary associated to a symplectic automorphism $S\in\Aut^{SO}\lr{\CL^{\perp}}=\Aut^{SO}\lr{\CL}$ we have
\begin{equation}
\lrq{\hat{U}_S, \Pi_{\CL}}=0.
\end{equation} 
This implies that the unnormalized ideal state
\begin{equation}
\ket{m}\coloneqq\Pi_{\CL}\ket{0}^{\otimes n} 
\end{equation}
is a $+1$ eigenvalue eigenstate of $\hat{U}_S$. $\ket{m}$ lives in the codespace of the GKP code and is a $+1$ eigenstate of the logical Clifford gate associated to $S$. For the single-qubit magic states in the following examples, a supply of these states together with logical Clifford gates and computational-basis measurements yields universal quantum computation~\cite{Bravyi_2005}.
Ref.~\cite{Bravyi_2005} distinguished between T- and H-types of magic states given by the single qubit Clifford orbit of the states \cite{Bravyi_2005}
\begin{align}
\ketbra{H}&=\frac{1}{2}\left(I+\frac{\hat{X}+\hat{Z}}{\sqrt{2}}\right),\\
\ketbra{T}&=\frac{1}{2}\left(I+\frac{\hat{X}+\hat{Y}+\hat{Z}}{\sqrt{3}}\right),
\end{align} 

\medskip
\paragraph*{Example: $\CL=\sqrt{2}\Z^2$}
For the square GKP code we have already identified the logical Hadamard gate realized by $e^{-i\pi/2 \hat{n}}$ as the only nontrivial logical Clifford gate realizable using a passive Gaussian unitary.
Furthermore, in the code space, the $+1$ eigenstate of the Hadamard gate is unique, so that we obtain $\ket{m}\propto\ket{H+}$, the $+1$ eigenvalue eigenstate of the logical Hadamard gate.
This fact was observed in \cite{AllGaussian}, where it was also shown that performing quantum error correction allows for the production of those magic states.
\medskip
\paragraph*{Example: $\CL=\sqrt{2}A_2$}
The hexagonal GKP code admits a passive Clifford implementation via $\hat U_R=e^{i\frac{2\pi}{3}\hat n}$~\cite{GCB}. Its projective Pauli action is that of $\hat H\hat P^{\dagger}$, as discussed in Sec.~\ref{sec:Cliffords}. To identify its eigenstates, we retain the signs: with $X=D(\bs e)$, $Z=D(\bs f)$, and $Y=iXZ$, the rotation acts as $X\mapsto -Y$, $Y\mapsto -Z$, and $Z\mapsto X$, and hence realizes $\hat X\hat H\hat P^{\dagger}$ up to an overall phase. Its eigenstates are Clifford-equivalent to the $\ket T$-type magic states defined in Ref.~\cite{Bravyi_2005}. Thus the state $\ket{m}$ obtained by formally projecting the vacuum onto the code space again yields a magic state.
\medskip

\section{Möbius acrobatics}\label{App:Moebius}

To further illustrate the behaviour of the Möbius action on the upper half plane we illustrate how it can be used to derive the Iwasawa and Bloch--Messiah decompositions, depicted in fig.~\ref{fig:IwasawaBloch}. Our presentation is guided by the example presented in ref.~\cite{Conrad_lectures}. The main ingredient in this understanding is the transitivity of $\SL_2\lr{\R}$ on the upper half-plane $\hh=\SL_2(\R).i\cong\SL_2(\R)/\SO_2(\R)$.

To derive the Iwasawa decomposition, recognize that  an arbitrary $z \in \hh$ can be written as
\begin{equation}
z=x+iy=\begin{pmatrix}
1 & x \\  0 & 1
\end{pmatrix}.\begin{pmatrix}
\sqrt{y} & 0 \\  0 & 1/\sqrt{y}
\end{pmatrix} . i,
\end{equation}
where the squeeze ``pushes" the point $z_0=i$ upwards to $z=iy$ and the final shear moves it horizontally to $z=x+iy$. Since every point $z \in \hh$ can be described by this sequence of Möbius transformations and the upper half plane is in bijection with elements of $\Sp_2\lr{\R}$ up to right multiplication by a rotation, we can deduce that every matrix $S \in \Sp_2\lr{\R}$ can be written as $S=NAK$, where $N$ and $A$ are shears and squeezes as above and $K\in \Sp_2\lr{\R}\cap \SO_2\lr{\R}$ is a rotation.

Similarly, every point $z$ can also be expressed by a squeeze and rotation acting on $z_0=i$, which leads to the Bloch-Messiah decomposition $S=K_1 A K_2$, where $A$ is again a squeeze and $K_1, K_2$ $\in \Sp_2\lr{\R}\cap \SO_2\lr{\R}$. The steps are geometrically sketched in fig.~\ref{fig:IwasawaBloch}.

\begin{figure}
\includegraphics[width=\columnwidth]{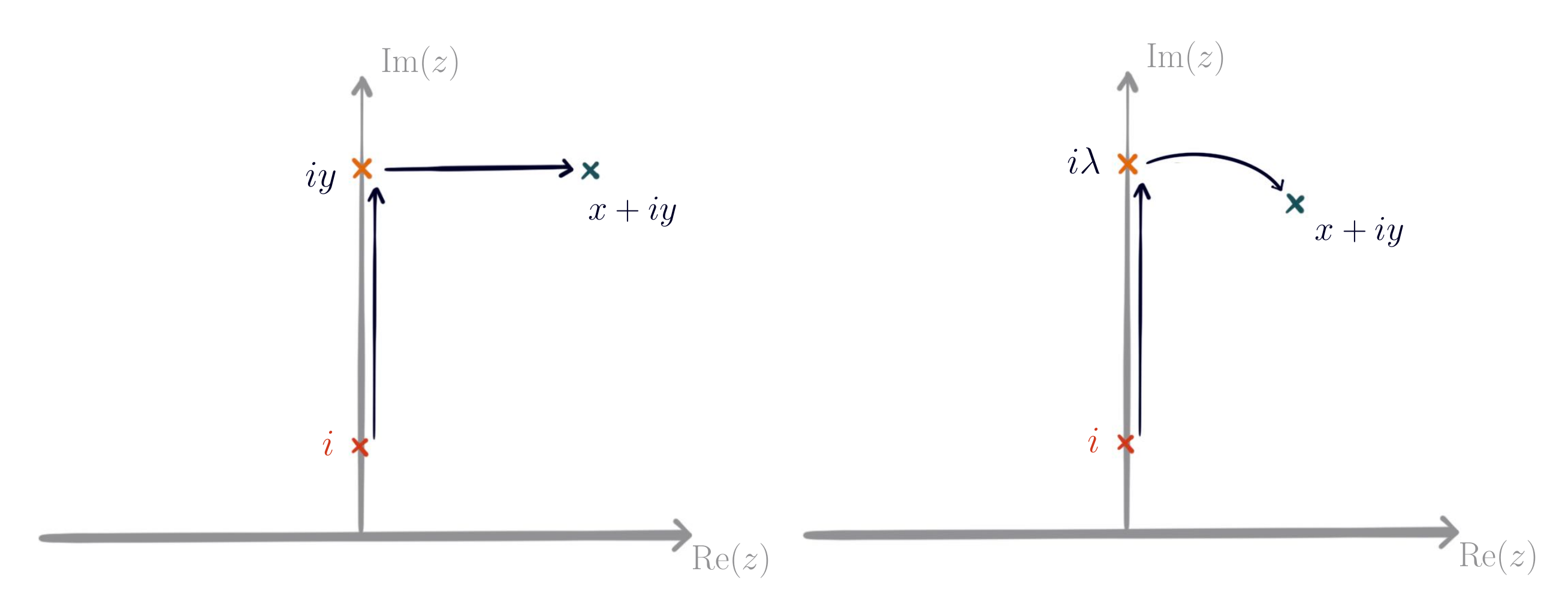}
\caption{We illustrate the Iwasawa (l.) and Bloch--Messiah (r.) decompositions of symplectic matrices. }
\label{fig:IwasawaBloch}
\end{figure}

\section{The Rademacher symbol for integral lifts}\label{app:Rademacher}

The Rademacher symbol is also given by \cite{Duke}
\begin{equation}
\psi\lr{A}=\Phi\lr{A}-3\mathrm{sign} \lr{c(a+d)}, 
\end{equation}
with $A=\begin{pmatrix}
a & b \\ c & d
\end{pmatrix} $
and where $\Phi\lr{A}$ is the Dedekind symbol 
\begin{align}
\Phi\lr{A}\coloneqq\begin{cases} \frac{b}{d} & \text{ if } c=0, \\
\frac{a+d}{c}-12\mathrm{sign}\lr{c} s(a,c), & \text{ if } c\neq 0,
\end{cases}
\end{align}
and the Dedekind sum $s(a,c)$ is given by
\begin{equation}
s(a,c)\coloneqq\sum_{n=1}^{|c|-1} \left(\!\left(\frac{n}{c}\right)\!\right)\left(\!\left(\frac{na}{c}\right)\!\right),
\end{equation}
where $\left(\!\left(x\right)\!\right)=x-\lfloor x \rfloor -1/2$ if $x$ is non-integer and $0$ otherwise. Since $ad\equiv 1 \bmod c$, one has $s(a,c)=s(d,c)$, so this agrees with the usual convention $\Phi(A)=\frac{a+d}{c}-12\,\mathrm{sign}(c)\,s(d,c)$.
\end{document}